\documentclass[aps,showpacs]{revtex4}

\usepackage{amsfonts}
\usepackage{amssymb}
\usepackage{amsmath}
\usepackage{dcolumn}
\usepackage{graphicx}
\usepackage{bm}
\usepackage{wasysym}

\begin{document}

\title{Quantum spin models for the SU($n$)$_1$ Wess-Zumino-Witten model}
\author{Hong-Hao Tu,$^{1}$ Anne E. B. Nielsen,$^{1}$ and Germ{\'a}n Sierra$%
^{2}$}
\affiliation{$^{1}$Max-Planck-Institut f\"ur Quantenoptik, Hans-Kopfermann-Str.\ 1,
D-85748 Garching, Germany\\
$^{2}$Instituto de F\'isica Te\'orica, UAM-CSIC, Madrid, Spain}

\begin{abstract}
We propose 1D and 2D lattice wave functions constructed from the SU($n$)$%
_{1} $ Wess-Zumino-Witten (WZW) model and derive their parent Hamiltonians.
When all spins in the lattice transform under SU($n$) fundamental
representations, we obtain a two-body Hamiltonian in 1D, including the SU($n$%
) Haldane-Shastry model as a special case. In 2D, we show that the wave
function converges to a class of Halperin's multilayer fractional quantum
Hall states and belongs to chiral spin liquids. Our result reveals a hidden
SU($n$) symmetry for this class of Halperin states. When the spins sit on
bipartite lattices with alternating fundamental and conjugate
representations, we provide numerical evidence that the state in 1D exhibits
quantum criticality deviating from the expected behaviors of the SU($n$)$_{1}$
WZW model, while in 2D they are chiral spin liquids being consistent with
the prediction of the SU($n$)$_{1}$ WZW model.
\end{abstract}

\maketitle

\section{Introduction}

For decades, SU($n$) quantum antiferromagnets have been an extensively
studied class of strongly correlated systems in condensed matter. Initially,
an important motivation of studying these models is that they may shed light
on the properties of the spin-1/2 antiferromagnetic Heisenberg models with
SU(2) symmetry \cite{affleck1988b,marston1989,read1989,read1990}, which are
relevant for many strongly correlated electronic materials, including
undoped high-$T_{c}$ superconductors. Similar to the large-$n$ expansion
used in quantum chromodynamics, generalizing SU(2)-symmetric models to SU($n$%
)-symmetric models allows stable mean-field solutions in the large-$n$ limit
\cite{arovas1988,auerbach1994}, and furthermore, systematic calculations of
corrections (organized in powers of $1/n$) can be carried out. Later on, the
proposal \cite{li1998,yamashita1998} that the SU(4) Heisenberg model might
describe certain materials with coupled spin-orbital degrees of freedom \cite%
{kugel1973} brings SU($n$) models closer to physical reality. By now, more
and more evidences show that, depending on the magnitude of $n$, spatial
dimensionality, lattice geometry, and form of couplings, the SU($n$) models
can support a zoo of exotic quantum states of matter \cite%
{azaria1999,assaraf1999,bossche2001,zhang2001,toth2010,corboz2011,corboz2012a,corboz2012b,bauer2012,corboz2013}%
. Recently, considerable progress has been achieved in the experimental
study of multi-flavor cold atoms in optical lattices \cite%
{taie2010,taie2012,zhang2014,scazza2014}. With these experimental setups,
atom species, lattice geometries, and interaction strengths can be
manipulated and engineered in a highly controllable way \cite%
{bloch2008,cazalilla2014}. The experimental advance spurs further
theoretical investigations \cite%
{wu2003,honerkamp2004,cazalilla2009,gorshkov2010,xu2010,manmana2011,messio2012,nonne2013,cai2013}
of the SU($n$) physics in the context of cold atomic setups. One may expect
that, in the near future, the rich SU($n$) physics might be experimentally
explored in an unprecedented depth.

From the theoretical point of view, the SU($n$) models are notoriously hard
to tackle. Needless to say, the validity of the large-$n$ solutions is
questionable for physically relevant small $n$ cases. Moreover, the SU($n$)
models usually suffer from the sign problem in quantum Monte Carlo
simulations (except for special cases \cite%
{frischmuth1999,harada2003,kawashima2007,kaul2014}), making them very
difficult even for numerical study. For these models, important insights are
gained from very few exactly solvable models, including integrable models
and AKLT-type models. The former ones are restricted to 1D, including e.g.\
the SU($n$) Uimin-Lai-Sutherland (ULS) model \cite%
{uimin1970,lai1974,sutherland1975} and the SU($n$) generalization \cite%
{kawakami1992,ha1992,kiwata1992} of the spin-1/2 Haldane-Shastry (HS) model
\cite{haldane1988,shastry1988}, both of which exhibit Tomonaga-Luttinger
liquid behaviors. The SU($n$) AKLT-type models \cite%
{affleck1991,chen2005,greiter2007a,greiter2007b,arovas2008,orus2011}
generalize the SU(2) AKLT models \cite{affleck1987,affleck1988a}, by
extending the SU($n$) singlets over multiple sites,\ and can be defined in one and higher dimensions.

Recently, a new approach of proposing strongly correlated wave functions has
been suggested in Refs.~\cite{cirac2010,nielsen2011,nielsen2012}. This
approach generalizes Moore and Read's construction \cite{moore1991} of
fractional quantum Hall (FQH) wave functions in the continuum, by expressing
both 1D and 2D \textit{lattice} wave functions as chiral correlators of
conformal field theories (CFTs). Apart from that, for rational CFTs, the
existence of null fields allows to derive a (long-range) parent Hamiltonian
\cite{nielsen2011}. Following this approach, wave functions have been
constructed for the SU(2)$_{k}$\ and SO($n$)$_{1}$\ WZW models \cite%
{nielsen2011,tu2013}, as well as $c=1$ free boson CFTs at particular
rational radii \cite{tu2014}. These simple wave functions, together with
their parent Hamiltonians, provide important insight into the properties of
their corresponding short-range realistic Hamiltonians \cite{nielsen2013},
which are hard to solve directly.

In this work, we construct spin wave functions using the SU($n$)$_{1}$ WZW
model and derive parent Hamiltonians of these states in 1D and 2D. In
particular, we focus on two cases: 1) lattices with all spins transforming
under SU($n$) \textit{fundamental} representations and 2) lattices with a
mixture of SU($n$) \textit{fundamental} and \textit{conjugate}
representations. In the former case, when the lattice sites are sitting on a
unit circle in the complex plane, we derive a two-body parent Hamiltonian.
This Hamiltonian can be viewed as an inhomogeneous extension of the SU($n$)
HS model. It recovers the SU($n$) HS model when the lattice sites are
\textit{uniformly} distributed on the unit circle, which we call 1D uniform
case. In 2D, we find that, on an infinite plane, the wave function converges
to a special class of Halperin states that appeared in the context of the
multilayer FQH effect. Interestingly, this reveals a hidden SU($n$) symmetry
for this class of Halperin states. Further numerical calculations based on
topological entanglement entropy (TEE) \cite{kitaev2006,levin2006} agree
with the prediction from the SU($n$)$_{1}$ WZW model and confirm that these
2D states are chiral spin liquids \cite{kalmeyer1987,kalmeyer1989,wen1989}.
For the more general case of wave functions with both fundamental and
conjugate representations, we concentrate on \textit{bipartite} lattices
with alternating fundamental and conjugate representations. In 1D uniform
case, the wave function exhibits logarithmically increasing entanglement
entropy and powerlaw decaying correlation functions, indicating quantum
critical behaviors. Surprisingly, the estimated central charges for $n=3$
and $4$ show clear deviations from the expected values for the SU($n$)$_{1}$
WZW model. In 2D, we find that the states are again chiral spin liquids,
with TEE being consistent with the prediction of the SU($n$)$_{1}$ WZW model.

The paper is organized as follows. In Sec.~\ref{sec:sun}, we explain how we
construct wave functions of spin systems from primary fields of the SU($n$)$%
_1$ WZW models, and we derive decoupling equations that form the basis for
obtaining parent Hamiltonians of the states. In Sec.~\ref{sec:FF}, we
consider the wave functions obtained from primary fields that transform
under the fundamental representation of SU($n$). We provide explicit
analytical expressions for the wave functions and compute the TEE of the
states in 2D numerically. In Sec.~\ref{sec:hamFF}, we derive parent
Hamiltonians of the states constructed from the fundamental representation.
For a uniform lattice in 1D this Hamiltonian reduces effectively to the SU($%
n $) HS model, and we also discuss CFT predictions for the spectrum of this
model. In Sec.~\ref{sec:FC}, we consider the more general case of wave
functions constructed from primary fields transforming either under the
fundamental or the conjugate representation of SU($n$). The wave functions
are expressed analytically, and we investigate their properties through
Monte Carlo simulations. Parent Hamiltonians of the states are derived in
Sec.~\ref{sec:hamFC}, where we also discuss possibilities for obtaining a
truncated short-range version of the Hamiltonian. Finally, Sec.~\ref%
{sec:conclusion} concludes the paper.

\section{Constructing quantum spin models from the SU($n$)$_{1}$ WZW model}

\label{sec:sun}

\subsection{Wave functions}

Before constructing the wave functions, let us briefly review the SU($n$)$%
_{1}$ WZW model \cite{francesco1997}. This rational CFT has $n$ primary
fields, denoted by $\Lambda _{a}$, with $a=0,1,\ldots ,n-1$, corresponding
to particular SU($n$) irreducible representations. The primary field $%
\Lambda _{0}$ is an SU($n$) singlet, which is also the identity field with
conformal weight $h(\Lambda _{0})=0$. The next primary field $\Lambda _{1}$
is the SU($n$) fundamental representation, corresponding to a single box
when the SU($n$) irreducible representations are represented as the Young
tableaux. In general, the primary field $\Lambda _{a}$ corresponds to a
Young tableau with a single column and $a$ rows. Accordingly, $\Lambda _{a}$
consists of $\dim \Lambda _{a}=\binom{n}{a}$ components, and we write these
components as $\Lambda_{a,\alpha}$, where $\alpha\in\{1,2,\ldots,\dim%
\Lambda_a\}$.

The central charge $c$, conformal weights $h(\Lambda _{a})$, and fusion
rules of the SU($n$)$_{1}$ WZW model are given by \cite{bouwknegt1999}%
\begin{equation}
c=n-1\text{, \ \ \ \ \ }h(\Lambda _{a})=\frac{a(n-a)}{2n},\text{ \ \ \ \ \ }%
\Lambda _{a}\otimes \Lambda _{b}=\Lambda _{a+b\text{ }(\text{\textrm{mod} }%
n)}.  \label{eq:SUnWZW}
\end{equation}%
As we shall discuss further below, the SU($n$)$_{1}$ WZW model has a
free-field representation with $n-1$ free bosons. In this representation,
the primary fields are conveniently realized using vertex operators.

To build lattice wave functions, we consider $N_\text{T}$ spins sitting at
the fixed positions $z_{j}$ ($j=1,2,\ldots ,N_{\text{T}}$) in the complex
plane. Following Ref.~\cite{nielsen2011}, we define lattice wave functions
\begin{equation}
|\Psi\rangle =\sum_{\alpha_1,\alpha_2,\ldots,\alpha_{N_{\text{T}}}} \langle
0|\Lambda_{a_{1},\alpha_1}(z_{1}) \Lambda_{a_{2},\alpha_2}(z_{2})\ldots
\Lambda_{a_{N_{\text{T}}},\alpha_{N_{\text{T}}}}(z_{N_{\text{T}}})|0\rangle
|\alpha_1,\alpha_2,\ldots,\alpha_{N_{\text{T}}}\rangle
\label{eq:wavefunction}
\end{equation}%
that are chiral correlators of primary fields. Here, $|0\rangle $ is the
vacuum of the CFT and $|\alpha_j\rangle$ are the basis vectors of the
internal state of spin number $j$. CFT states of the form (\ref%
{eq:wavefunction}) can be seen as a special type of matrix product states in
which the finite-dimensional matrices have been replaced by infinite-dimensional conformal fields. They are therefore sometimes referred to as
infinite-dimensional-matrix product states (IDMPS).

Regarding the wave function (\ref{eq:wavefunction}), there are several
comments in order. First, choosing the primary field $\Lambda _{a_{j}}$ at
site $j$ requires that the spin at this site also transforms under the SU($n$%
) irreducible representation corresponding to a Young tableau with one
column and $a_{j}$ rows. Note that the SU($n$)$_{1}$ WZW model does not have
primary fields corresponding to a Young tableaux with more than one column.
Secondly, the fusion rules in (\ref{eq:SUnWZW}) always have a \textit{unique}
fusion outcome, which ensures that the wave function (\ref{eq:wavefunction})
is a unique function. Lastly, to have a nonvanishing wave function, the $N_{%
\text{T}}$ primary fields in (\ref{eq:wavefunction}) must fuse into the
identity $\Lambda_{0}$ (i.e.\ the SU($n$) singlet),
\begin{equation}  \label{eq:fusion}
\Lambda _{a_{1}}\otimes \Lambda _{a_{2}}\otimes \cdots \otimes \Lambda
_{a_{N_{\text{T}}}}=\Lambda _{0}.
\end{equation}

In this work, we shall focus on the case, where each of the spins belong
either to the SU($n$) \textit{fundamental} representation $\Lambda_{1}$ or
to the SU($n$) \textit{conjugate} representation $\Lambda_{n-1}$. We shall
denote the sublattice of spins transforming under the fundamental
(conjugate) representation by $A$ ($B$),
\begin{eqnarray}
&&A:\text{ Fundamental representation},  \notag \\
&&B:\text{ Conjugate representation},
\end{eqnarray}
and we shall let $N$ ($\bar{N}$) denote the number of spins in $A$ ($B$)
such that $N+\bar{N}=N_{\text{T}}$. The condition (\ref{eq:fusion}) then
gives that $(N-\bar{N})/n$ must be an integer, and we shall assume this to
be the case throughout. Note that the fundamental and conjugate
representations are the same for $n=2$, so that there is only one state in
this particular case. For $n\geq3$, however, they are different.

Before we continue with the above case, let us note that other choices for
the primary fields are possible. For instance, for even $n$, one could use
the primary field $\Lambda _{n/2}$ (self-conjugate representation) to build
the wave function, according to the fusion rule $\Lambda _{n/2}\otimes
\Lambda _{n/2}\otimes \cdots \otimes \Lambda _{n/2}=\Lambda _{0}$ ($N_{\text{%
T}}$ even). For the SU(4) case, one has SU(4)$_{1}\simeq $ SO(6)$_{1}$ and
the SU(4) self-conjugate primary field $\Lambda_{2}$ becomes the vector
representation of SO(6) with conformal weight $h(\Lambda _{2})=1/2$, which
can be interpreted as a Majorana field and has been considered in Ref.~\cite%
{tu2013}. Although we only consider states constructed from the fundamental
and conjugate representations below, we note that the formalism we develop
is general and that other cases can be treated in a similar way.

In the following, we shall find it convenient to use the notation
\begin{equation}
\varphi _{\alpha _{j}}(z_{j})=\left\{
\begin{array}{cl}
\Lambda _{1,\alpha _{j}}(z_{j}) & \text{for }j\in A \\
\Lambda _{n-1,\alpha _{j}}(z_{j}) & \text{for }j\in B%
\end{array}%
\right. .
\end{equation}%
We can then write the wave functions that we are interested in as
\begin{equation}
|\Psi \rangle =\sum_{\alpha _{1},\ldots ,\alpha _{N_{\text{T}}}=1}^{n}\Psi
(\alpha _{1},\ldots ,\alpha _{N_{\text{T}}})|\alpha _{1},\ldots ,\alpha _{N_{%
\text{T}}}\rangle ,  \label{eq:wf}
\end{equation}%
where
\begin{equation}
\Psi (\alpha _{1},\ldots ,\alpha _{N_{\text{T}}})=\langle 0|\varphi _{\alpha
_{1}}(z_{1})\varphi _{\alpha _{2}}(z_{2})\cdots \varphi _{\alpha _{N_{\text{T%
}}}}(z_{N_{\text{T}}})|0\rangle .  \label{eq:wavefunction1}
\end{equation}%
Since we shall often refer to the wave function, for which all the primary
fields belong to the fundamental representation, we shall give this wave
function a particular name: $|\Psi _{\mathrm{F}}\rangle $. Explicit
representations of $|\Psi _{\mathrm{F}}\rangle $ and $|\Psi \rangle $ will
be discussed in Secs.~\ref{sec:FF} and \ref{sec:FC}, respectively. In the
next two subsections, we shall use their abstract forms to derive relevant
null fields and their corresponding decoupling equations, which are our
starting point for deriving parent Hamiltonians.

\subsection{Null vectors}

As a rational CFT, the SU($n$)$_{1}$ WZW model has null vectors in its Verma
modules of the Kac-Moody algebra. According to Ref.~\cite{nielsen2011},
identifying proper null vectors and deriving decoupling equations for the
chiral correlators are the key for constructing parent Hamiltonians of the
wave functions. In this subsection, we derive the null vectors relevant for (%
\ref{eq:wavefunction1}).

The SU($n$)$_{1}$ Kac-Moody algebra is defined by%
\begin{equation}
\lbrack J_{m}^{a},J_{m^{\prime }}^{b}]=if_{abc}J_{m+m^{\prime }}^{c}+\frac{m%
}{2}\delta _{ab}\delta _{m+m^{\prime },0},\text{ \ \ \ \ \ }m,m^{\prime }\in
\mathbb{Z},
\end{equation}%
where $J_{m}^{a}=\oint_0 \frac{dz}{2\pi i}z^{m}J^{a}(z)$ is the $m$th mode
of the Kac-Moody current $J^{a}(z)$ and $f_{abc}$ are the structure
constants of the SU($n$) Lie algebra. Here and later on, we shall always
assume that repeated indices are summed over. The operator product expansion
(OPE) between the Kac-Moody currents and a primary field is \cite%
{francesco1997}
\begin{equation}
J^{a}(z)\varphi _{\alpha }(w)\sim -\frac{1}{z-w}\sum_{\beta }(t^{a})_{\alpha
\beta }\varphi _{\beta }(w),  \label{eq:OPE}
\end{equation}%
where the matrices $t^{a}$ with elements $(t^{a})_{\alpha\beta}$ are the
generators of SU($n$) in the representation of the primary field. Let us
note here that the generators in the fundamental and conjugate
representations are related though a complex conjugation and a
multiplication by a minus sign, i.e.,
\begin{equation}  \label{eq:ta}
t^a=\left\{%
\begin{array}{cl}
\tau^a & \text{(fundamental representation)} \\
-(\tau^a)^* & \text{(conjugate representation)}%
\end{array}%
\right.,
\end{equation}
where $\tau^a$ are the generators in the fundamental representation (see
Appendix \ref{sec:appSUn}).

To the primary field $\varphi_\alpha(z)$, one associates a primary state $%
|\varphi_\alpha\rangle$ satisfying the following properties \cite%
{francesco1997}:
\begin{equation}
|\varphi_{\alpha}\rangle=\varphi_{\alpha}(0)|0\rangle,\text{ \ }%
J_{0}^{a}|\varphi_{\alpha}\rangle =-\sum_{\beta}(t^{a})_{\alpha \beta
}|\varphi_{\beta}\rangle ,\text{ \ }J_{n}^{a}|\varphi_{\alpha}\rangle=0,%
\text{ \ }n>0,  \label{eq:PrimayState}
\end{equation}%
and descendant states are obtained by multiplying $|\varphi_\alpha\rangle$
by any number of current operators $J_{n}^{a}$ with $n<0$. A null state is a
state that is at the same time a descendant and a primary state. Since the
wave function (\ref{eq:wavefunction1}) only involves primary fields
belonging to the fundamental or the conjugate representation, we shall here
only need to deal with the two Verma modules formed by the corresponding
primary states, as well as their descendants.

\begin{figure}[tbp]
\centering
\includegraphics[scale=0.6]{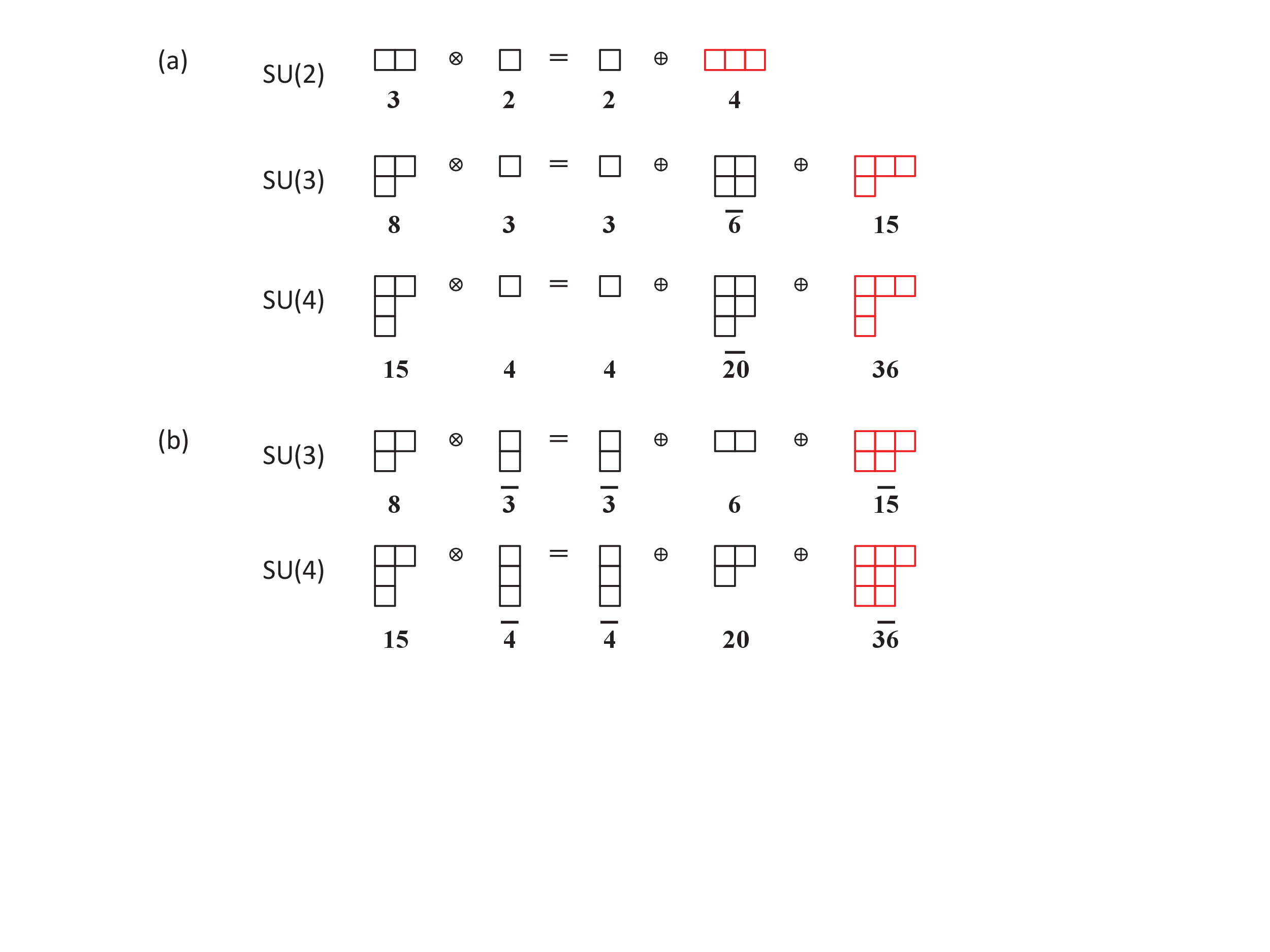}
\caption{Tensor product decompositions of the SU($n$) adjoint representation
and (a) the fundamental ((b) conjugate) representation with Young tableaux.
The null vectors belong to the SU($n$) high-dimensional representation (with
red color).}
\label{fig:NullVector}
\end{figure}

Let us first consider the primary field $\Lambda_{1,\alpha}(z)$ belonging to
the \textit{fundamental} representation. In Virasoro level $m=1$, we look
for null vectors with the following form:
\begin{equation}
|\chi ^{q}\rangle =\sum_{a,\alpha }W_{q,a\alpha
}J_{-1}^{a}|\Lambda_{1,\alpha}\rangle ,  \label{eq:NullVector}
\end{equation}%
where $W_{q,a\alpha }$ can be interpreted as Clebsch-Gordan coefficients
satisfying $\sum_{a\alpha }W_{q,a\alpha }^{\ast }W_{q^{\prime },a\alpha
}=\delta _{qq^{\prime }}$. They come from the tensor product decomposition
of the $(n^{2}-1)$-dimensional SU($n$) adjoint representation (carried by $%
J_{-1}^{a}$) and the fundamental representation (carried by $%
|\Lambda_{1,\alpha}\rangle $),%
\begin{equation}
(n^{2}-1)\otimes n=n\oplus \frac{1}{2}n(n+1)(n-2)\oplus \frac{1}{2}%
n(n-1)(n+2),  \label{eq:TensorDecomposition}
\end{equation}%
where the irreducible representations are denoted by their dimensions (they
are not distinguished with their complex conjugate representations). Fig.~%
\ref{fig:NullVector}(a) shows the tensor product decomposition (\ref%
{eq:TensorDecomposition}) for $n=2,3,4$, using the Young tableaux. We have
found that, for SU($n$)$_{1}$ WZW model with all $n$, null vectors indeed
exist in Virasoro level $m=1$, and they belong to the SU($n$) representation
with dimension $\frac{1}{2}n(n-1)(n+2)$ in (\ref{eq:TensorDecomposition}).
In practice, the Clebsch-Gordan coefficients $W_{q,a\alpha }$ in (\ref%
{eq:NullVector})\ can be determined by requiring the null vector condition $%
\langle \chi ^{b^{\prime }}|\chi ^{b}\rangle =0$.

For our purpose, we redefine the null vectors as \cite{nielsen2011}%
\begin{eqnarray}
|\chi ^{a,\alpha }\rangle &=&\sum_{q}W_{q,a\alpha }^{\ast }|\chi ^{q}\rangle
\notag \\
&=&\sum_{b,\beta }(K_{\mathrm{F}})_{b\beta }^{a\alpha }J_{-1}^{b}|\Lambda
_{1,\beta }\rangle ,  \label{eq:NullVector1}
\end{eqnarray}%
where $(K_{\mathrm{F}})_{b\beta }^{a\alpha }$ is given by%
\begin{equation}
(K_{\mathrm{F}})_{b\beta }^{a\alpha }=\sum_{q}W_{q,a\alpha }^{\ast
}W_{q,b\beta }.
\end{equation}%
$\mathbf{K}_{\mathrm{F}}$ can be viewed as a matrix with its entries being $(%
\mathbf{K}_{\mathrm{F}})_{a\alpha ,b\beta }=(K_{\mathrm{F}})_{b\beta
}^{a\alpha }$, and it is a projector (i.e. $\mathbf{K}_{\mathrm{F}}^{2}=%
\mathbf{K}_{\mathrm{F}}$) onto the SU($n$) irreducible representation with
dimension $\frac{1}{2}n(n-1)(n+2)$. This also lead to an additional
equation, $\sum_{a}t^{a}(K_{\mathrm{F}})_{b}^{a}=0$, where $(K_{\mathrm{F}%
})_{b}^{a}$ is a matrix with entries $[(K_{\mathrm{F}})_{b}^{a}]_{\alpha
\beta }=(K_{\mathrm{F}})_{b\beta }^{a\alpha }$. These two equations are
sufficient for determining the explicit form of $(K_{\mathrm{F}})_{b}^{a}$.
For general $n$, we obtain%
\begin{equation}
(K_{\mathrm{F}})_{b}^{a}=\frac{n+2}{2(n+1)}\delta _{ab}+\frac{1}{2(n+1)}%
[nd_{abc}-(n+2)if_{abc}]t^{c},
\end{equation}%
where $d_{abc}$ is a totally symmetric tensor (see Appendix \ref{sec:appSUn}%
).

If we build the null vector at Virasoro level $m=1$ using the primary state $%
|\Lambda _{n-1,\alpha }\rangle $ in the conjugate representation, the
representations appearing in (\ref{eq:TensorDecomposition}) would be their
complex conjugate representations. See Fig.~\ref{fig:NullVector}(b) for this
tensor product decomposition for $n=3$ and $4$. As a result, the
Clebsch-Gordan coefficients in (\ref{eq:NullVector}) for obtaining the null
vectors would be their complex conjugate. Then, the corresponding null
vectors can be written as%
\begin{equation}
|\chi ^{a,\alpha }\rangle =\sum_{b,\beta }(K_{\mathrm{C}})_{b\beta
}^{a\alpha }J_{-1}^{b}|\Lambda _{n-1,\beta }\rangle ,  \label{eq:NullVector2}
\end{equation}%
where $(K_{\mathrm{C}})_{b\beta }^{a\alpha }=(K^{\ast }_{\mathrm{F}%
})_{b\beta }^{a\alpha }$.

Utilizing (\ref{eq:ta}), we can combine (\ref{eq:NullVector1}) and (\ref%
{eq:NullVector2}) into a single expression
\begin{equation}
|\chi ^{a,\alpha }\rangle =\sum_{b,\beta }K_{b\beta }^{a\alpha
}J_{-1}^{b}|\varphi_{\beta}\rangle ,  \label{eq:NullVector12}
\end{equation}
where $K_{b\beta }^{a\alpha }$ are the matrix elements of
\begin{equation}
K_{b}^{a}=\frac{n+2}{2(n+1)}\delta _{ab}+\frac{1}{2(n+1)}%
[nrd_{abc}-(n+2)if_{abc}]t^{c}.  \label{eq:KMatrix}
\end{equation}%
Here, $r=+1$ for the fundamental representation, $r=-1$ for the conjugate
representation, and $t^{c}$ are the generators in the considered
representation.

\subsection{Decoupling equations}

Following Ref.~\cite{nielsen2011}, a set of decoupling equations can be
derived for the chiral correlator (\ref{eq:wavefunction1}) using the null
vectors (\ref{eq:NullVector12}). These decoupling equations provide
operators annihilating the wave functions, which can be used to build parent
Hamiltonians.

The null state (\ref{eq:NullVector12}) corresponds to the following null
field:%
\begin{equation}
\chi ^{a,\alpha }(z_{i})=\oint_{z_{i}}\frac{dz}{2\pi i}\frac{1}{z-z_{i}}%
\sum_{b,\beta }K_{b\beta }^{a\alpha }J^{b}(z)\varphi _{\beta }(z_{i}).
\label{eq:NullField}
\end{equation}%
By definition of the null field, substituting it into the wave function (\ref%
{eq:wavefunction1}), one obtains a vanishing expression%
\begin{eqnarray}
0 &=&\sum_{\alpha _{1},\ldots ,\alpha _{N}}\langle \varphi _{\alpha
_{1}}(z_{1})\cdots \chi ^{a,\alpha _{i}}(z_{i})\cdots \varphi _{\alpha
_{N}}(z_{N})\rangle |\alpha _{1},\ldots ,\alpha _{N}\rangle \text{ \ }%
\forall a  \notag \\
&=&\sum_{\alpha _{1},\ldots ,\alpha _{N}}\sum_{b,\beta _{i}}K_{b,\beta
_{i}}^{a,\alpha _{i}}\oint_{z_{i}}\frac{dz}{2\pi i}\frac{1}{z-z_{i}}\langle
\varphi _{\alpha _{1}}(z_{1})\cdots J^{b}(z)\varphi _{\beta
_{i}}(z_{i})\cdots \varphi _{\alpha _{N}}(z_{N})\rangle |\alpha _{1},\ldots
,\alpha _{N}\rangle .
\end{eqnarray}%
After deforming the integral contour and using the OPE (\ref{eq:OPE})
between the Kac-Moody currents and primary fields, we arrive at%
\begin{eqnarray}
0 &=&\sum_{\alpha _{1},\ldots ,\alpha _{N}}\sum_{j(\neq i)}\sum_{\alpha
_{j}^{\prime }} \frac{(t_{j}^{b})_{\alpha _{j}\alpha _{j}^{\prime }}}{%
z_{i}-z_{j}}\sum_{b,\beta _{i}}K_{b,\beta _{i}}^{a,\alpha _{i}}\langle
\varphi _{\alpha _{1}}(z_{1})\cdots \varphi _{\alpha _{j}^{\prime
}}(z_{j})\cdots \varphi _{\beta _{i}}(z_{i})\cdots \varphi _{\alpha
_{N}}(z_{N})\rangle |\alpha _{1},\ldots ,\alpha _{j},\ldots ,\alpha
_{i},\ldots ,\alpha _{N}\rangle  \notag \\
&=&\sum_{j(\neq i),b}\frac{t_{j}^{b}}{z_{i}-z_{j}}(K^{(i)})_{b}^{a}|\Psi%
\rangle \text{ \ }\forall a,
\end{eqnarray}%
where $(K^{(i)})_{b}^{a}$ denotes the operator $K_{b}^{a}$ in (\ref%
{eq:KMatrix}) acting on spin number $i$ and $(t_{j}^{b})_{\alpha _{j}\alpha
_{j}^{\prime }}$ denote the matrix elements of the operator $t^b$ acting on
spin number $j$. (Note that the representation chosen for $t_j^b$ is the
same as the representation of spin number $j$.) Thus, the resulting
decoupling equation yields a set of operators%
\begin{equation}
\mathcal{P}_{i}^{a}(z_{1},\ldots ,z_{N})=\sum_{j(\neq i),b}\frac{t_{j}^{b}}{%
z_{i}-z_{j}}(K^{(i)})_{b}^{a},
\end{equation}%
which annihilate the wave function $|\Psi\rangle$, i.e.\ $\mathcal{P}%
_{i}^{a}(z_{1},\ldots ,z_{N})|\Psi\rangle =0$ $\forall i,a$. Together with
the fact that $|\Psi\rangle $ is a global SU($n$) singlet, $%
T^{a}|\Psi\rangle=0$\ with $T^{a}=\sum_{i}t_{i}^{a}$, we obtain%
\begin{equation}
\mathcal{C}_{i}^{a}(z_{1},\ldots ,z_{N})|\Psi\rangle=0,
\end{equation}%
where $\mathcal{C}_{i}^{a}(z_{1},\ldots ,z_{N})$ is given by%
\begin{eqnarray}
\mathcal{C}_{i}^{a}(z_{1},\ldots ,z_{N}) &=&\sum_{j(\neq
i),b}w_{ij}(K^{(i)})_{b}^{a}t_{j}^{b}  \notag \\
&=&\frac{n+2}{2(n+1)}\sum_{j(\neq i)}w_{ij}[t_{j}^{a}+(\frac{n}{n+2}%
r_id_{abc}+if_{abc})t_{i}^{b}t_{j}^{c}]  \label{eq:Annihilator1}
\end{eqnarray}%
and $w_{ij}=(z_{i}+z_{j})/(z_{i}-z_{j})$. For SU(2), we have $d_{abc}=0$ and
(\ref{eq:Annihilator1}) recovers the result in Ref.~\cite{nielsen2011}.
Utilizing the formulas in Appendix \ref{sec:appSUn}, we get
\begin{equation}
\mathcal{C}_{i}^{a}(z_{1},\ldots ,z_{N})=\sum_{j(\neq i)}w_{ij}[\frac{1}{2}%
t_{j}^{a}-\frac{1}{n+1}t_{i}^{a}(\vec{t}_{i}\cdot \vec{t}_{j})+(\vec{t}%
_{i}\cdot \vec{t}_{j})t_{i}^{a}],  \label{eq:Annihilator2}
\end{equation}
which is a convenient form for constructing parent Hamiltonians.

\subsection{Vertex operator representation}

After working out the decoupling equations for (\ref{eq:wavefunction1})
using an abstract form of the primary fields, we now turn to an explicit
representation of these primary fields, using chiral vertex operators. This
is possible, since SU($n$)$_{1}$ WZW model is equivalent to a free theory of
$n-1$ massless bosons.

For our purpose, it is convenient to label the spin states in each site by
their weights (eigenvalues of the Cartan generators). The state $|\alpha
\rangle $, $\alpha \in \{1,2,\ldots ,n\}$, in the fundamental representation
is therefore characterized by $n-1$ quantum numbers, which we collect into
the vector $\vec{m}_{\alpha}$ given explicitly by
\begin{equation}
\begin{array}{rcccccccccccl}
\vec{m}_{1} & = & \Big( & \frac{1}{2} & , & \frac{1}{2\sqrt{3}} & , & \ldots
& , & \frac{1}{\sqrt{2n(n-1)}} & \Big), &  &  \\
\vec{m}_{2} & = & \Big( & -\frac{1}{2} & , & \frac{1}{2\sqrt{3}} & , & \ldots
& , & \frac{1}{\sqrt{2n(n-1)}} & \Big), &  &  \\
\vec{m}_{3} & = & \Big( & 0 & , & -\frac{1}{\sqrt{3}} & , & \ldots & , &
\frac{1}{\sqrt{2n(n-1)}} & \Big), &  &  \\
& \vdots &  &  &  &  &  &  &  &  &  &  &  \\
\vec{m}_{n} & = & \Big( & 0 & , & 0 & , & \ldots & , & -\frac{n-1}{\sqrt{%
2n(n-1)}} & \Big). &  &
\end{array}
\label{eq:sunstates}
\end{equation}%
In the conjugate representation, the state $|\alpha\rangle $, $\alpha\in
\{1,2,\ldots ,n\}$, is characterized by the quantum numbers $-\vec{m}%
_{\alpha}$. The SU(3) and SU(4) weight diagrams are shown in Fig.~\ref%
{fig:SUnRep} as examples.

\begin{figure}[tbp]
\centering
\includegraphics[scale=0.6]{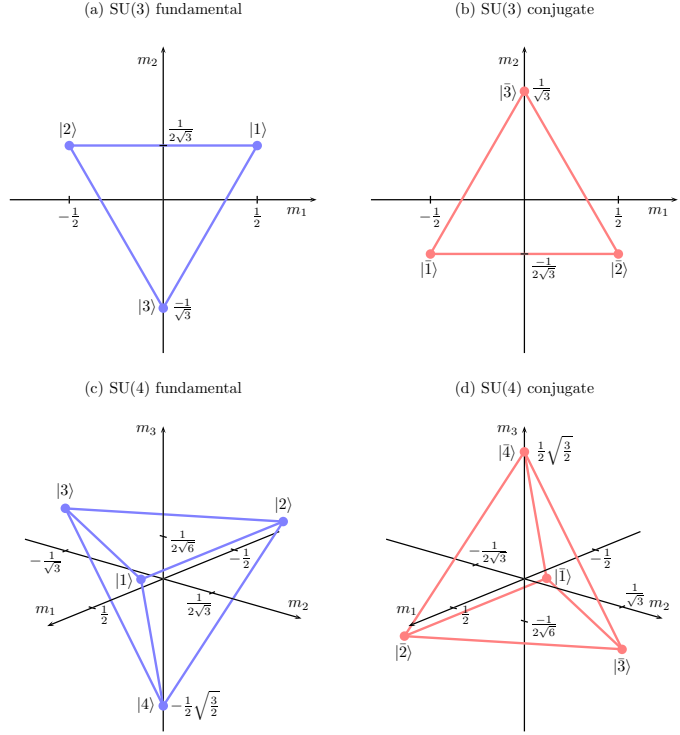}
\caption{Weight diagrams of the fundamental and conjugate representations
for SU(3) and SU(4). Here, $(m_1,m_2,m_3)$ is shorthand notation for the
components of the vectors $\vec{m}_\protect\protect\alpha$ and $-\vec{m}%
_\protect\protect\alpha$, respectively.}
\label{fig:SUnRep}
\end{figure}

Using the weights, the primary field $\varphi_{\alpha}(z)$ can be expressed
as
\begin{equation}  \label{eq:iMPSPrimaryField}
\varphi _{\alpha}(z) = \kappa _{\alpha}:\exp \left( i\sqrt{2}r\vec{m}%
_{\alpha}\cdot \vec{\phi}(z)\right) :{},
\end{equation}%
where $r=+1$ for the fundamental representation and $r=-1$ for the conjugate
representation as above. The colons denote normal ordering and $\vec{\phi}%
(z) $ is a vector of $n-1$ independent fields of free, massless bosons. The
factor $\kappa _{\alpha }$ is a Klein factor, commuting with the vertex
operators and satisfying Majorana-like anticommutation relations%
\begin{equation}  \label{eq:KleinFactor}
\{\kappa _{\alpha },\kappa _{\alpha ^{\prime }}\}=2\delta _{\alpha \alpha
^{\prime }}.
\end{equation}
Note that $\kappa_\alpha$ is the same in the fundamental and in the
conjugate representation. At this moment, the meaning of these Klein factors
is not clear. In fact, their role is to ensure that the wave function (\ref%
{eq:wavefunction1}) is an SU($n$) singlet. We will go back to this point
when discussing the wave functions in Sec.~\ref{sec:FF} and Sec.~\ref{sec:FC}%
.

Let us note that the vertex operators in (\ref{eq:iMPSPrimaryField}) have
the anticipated conformal weights, since
\begin{equation}
\vec{m}_{\alpha }\cdot \vec{m}_{\alpha }=h(\Lambda _{1})=h(\Lambda _{n-1})=%
\frac{n-1}{2n}.
\end{equation}%
Another quantity, which will be used in later sections, is $\vec{m}_{\alpha
}\cdot \vec{m}_{\alpha ^{\prime }}$ with $\alpha \neq \alpha ^{\prime }$. It
is easy to convince ourselves that this value does not depend on the states
we choose. For $\alpha \neq \alpha ^{\prime }$, we find
\begin{equation}
\vec{m}_{\alpha }\cdot \vec{m}_{\alpha ^{\prime }}=-\frac{1}{2n}.
\end{equation}%
Altogether, we thus conclude
\begin{equation}
\vec{m}_{\alpha }\cdot \vec{m}_{\alpha ^{\prime }}=\frac{1}{2}\delta
_{\alpha \alpha ^{\prime }}-\frac{1}{2n}.  \label{eq:mdotm}
\end{equation}

\section{Quantum states from the fundamental representation of SU($n$)}

\label{sec:FF}

In this section, we analyze the wave function (\ref{eq:wavefunction1}) in
detail, both theoretically and numerically, for the case where all spins
transform under the fundamental representation. First, the chiral correlator
can be evaluated and expressed in terms of a product of Jastrow factors \cite%
{francesco1997}
\begin{equation}
\Psi _{\mathrm{F}}(\alpha _{1},\alpha _{2},\ldots ,\alpha _{N})=\chi (\alpha
_{1},\alpha _{2},\ldots ,\alpha _{N})\delta _{\sum_{i}\vec{m}_{\alpha
_{i}}=0}\prod_{i<j}(z_{i}-z_{j})^{2\vec{m}_{\alpha _{i}}\cdot \vec{m}%
_{\alpha _{j}}},  \label{eq:Jastrow}
\end{equation}%
where $\chi (\alpha _{1},\alpha _{2},\ldots ,\alpha _{N})=\kappa _{\alpha
_{1}}\kappa _{\alpha _{2}}\cdots \kappa _{\alpha _{N}}$ is a $z_{j}$%
-independent phase factor to be determined below and the Kronecker delta
function $\delta _{\sum_{i}\vec{m}_{\alpha _{i}}=0}$, which is $1$ for $%
\sum_{i}\vec{m}_{\alpha _{i}}=0$ and zero otherwise, ensures charge
neutrality. Referring to Eq.~(\ref{eq:sunstates}), we observe that the
charge neutrality forces the number of spins $N_{\alpha }$ in the state $%
|\alpha \rangle $ to fulfill $N_{1}=N_{2}=\ldots =N_{n}$. This gives $%
N_{\alpha }=N/n$ for all $\alpha $, and we shall therefore assume $N/n$ to
be an integer whenever we consider states constructed from only the
fundamental representation of SU($n$). Utilizing (\ref{eq:mdotm}), we note
that (\ref{eq:Jastrow}) simplifies to
\begin{equation}
\Psi _{\mathrm{F}}(\alpha _{1},\alpha _{2},\ldots ,\alpha _{N})\propto \chi
(\alpha _{1},\alpha _{2},\ldots ,\alpha _{N})\delta _{\sum_{i}\vec{m}%
_{\alpha _{i}}=0}\prod_{i<j}(z_{i}-z_{j})^{\delta _{\alpha _{i}\alpha _{j}}}.
\label{eq:wfSim}
\end{equation}

We shall also find it useful to express the state $|\Psi _{\mathrm{F}%
}\rangle $ in another notation. For a given spin configuration $|\alpha
_{1},\alpha _{2},\ldots ,\alpha _{N}\rangle $, let $x_{j}^{(\alpha )}$,
where $j=1,2,\ldots ,N/n$, be the position within the ket of the $j$th spin
in the state $|\alpha \rangle $. For example, if we choose $n=3$ and $N=9$
and consider the state ket $|1,2,1,3,2,3,3,2,1\rangle $, we would have $%
x_{1}^{(1)}=1$, $x_{2}^{(1)}=3$, $x_{3}^{(1)}=9$, $x_{1}^{(2)}=2$, $%
x_{2}^{(2)}=5$, $x_{3}^{(2)}=8$, $x_{1}^{(3)}=4$, $x_{2}^{(3)}=6$, and $%
x_{3}^{(3)}=7$. We shall also write $\{x_{1\rightarrow \frac{N}{n}}^{(\alpha
)}\}$ or simply $\{x^{(\alpha )}\}$ as shorthand notation for $%
x_{1}^{(\alpha )},x_{2}^{(\alpha )},\ldots ,x_{\frac{N}{n}}^{(\alpha )}$. We
can then express $|\Psi _{\mathrm{F}}\rangle $ as
\begin{equation}
|\Psi _{\mathrm{F}}\rangle =\sum_{\{x^{(1)}\},\{x^{(2)}\},\ldots
,\{x^{(n)}\}\in S_{N}}\Psi _{\mathrm{F}}(\{x^{(1)}\},\{x^{(2)}\},\ldots
,\{x^{(n)}\})|\{x^{(1)}\},\{x^{(2)}\},\ldots ,\{x^{(n)}\}\rangle ,
\label{eq:wf1}
\end{equation}%
where $S_{N}$ is the symmetric group over the elements $\{1,2,\ldots ,N\}$
and
\begin{equation}
\Psi _{\mathrm{F}}(\{x^{(1)}\},\{x^{(2)}\},\ldots ,\{x^{(n)}\})\propto \chi
(\{x^{(1)}\},\{x^{(2)}\},\ldots ,\{x^{(n)}\})\prod_{\alpha
=1}^{n}\prod_{1\leq i<j\leq \frac{N}{n}}(z_{x_{i}^{(\alpha
)}}-z_{x_{j}^{(\alpha )}}).  \label{eq:wf2}
\end{equation}

Let us next determine $\chi $ from the condition that $|\Psi_{\mathrm{F}%
}\rangle$ should be an SU($n$) singlet. We shall find below that the wave
function $|\Psi _{\mathrm{F}}\rangle $ is proportional to the ground state
of the SU($n$) HS model if we choose $z_{j}=e^{2\pi ij/N}$ and
\begin{equation}
\chi =\mathrm{sgn}(x_{1}^{(1)},\ldots ,x_{N/n}^{(1)},x_{1}^{(2)},\ldots
,x_{N/n}^{(2)},\ldots ,x_{1}^{(n)},\ldots ,x_{N/n}^{(n)}),  \label{eq:chi}
\end{equation}%
where the right-hand side of (\ref{eq:chi}) is the sign of the permutation
needed to transform $x_{1}^{(1)},\ldots ,x_{N/n}^{(1)},x_{1}^{(2)},\ldots
,x_{N/n}^{(2)},\ldots ,x_{1}^{(n)},\ldots ,x_{N/n}^{(n)}$ into $1,2,\ldots
,N $. Since the ground state of the SU($n$) HS model is an SU($n$) singlet,
it follows that (\ref{eq:chi}) is the correct choice of $\chi $ for all
choices of $z_{j}$. The result (\ref{eq:chi}) can be obtained from $\chi
=\kappa _{\alpha _{1}}\kappa _{\alpha _{2}}\cdots \kappa _{\alpha _{N}}$ by
choosing the factors $\kappa _{\alpha }$ to be Klein factors, which satisfy
the Majorana-like anticommutation relation (\ref{eq:KleinFactor}), and
choosing to work in a sector, in which $\kappa _{1}\kappa _{2}\cdots \kappa
_{n}=1$. This follows from
\begin{eqnarray}
\kappa _{\alpha _{1}}\kappa _{\alpha _{2}}\cdots \kappa _{\alpha _{N}} &=&%
\mathrm{sgn}(x_{1}^{(1)},\ldots ,x_{N/n}^{(1)},x_{1}^{(2)},\ldots
,x_{N/n}^{(2)},\ldots ,x_{1}^{(n)},\ldots ,x_{N/n}^{(n)})\kappa
_{1}^{N/n}\kappa _{2}^{N/n}\cdots \kappa _{n}^{N/n}  \notag \\
&=&\left\{
\begin{array}{cc}
\mathrm{sgn}(x_{1}^{(1)},\ldots ,x_{N/n}^{(1)},x_{1}^{(2)},\ldots
,x_{N/n}^{(2)},\ldots ,x_{1}^{(n)},\ldots ,x_{N/n}^{(n)}) & \text{for }N/n%
\text{ even} \\
\mathrm{sgn}(x_{1}^{(1)},\ldots ,x_{N/n}^{(1)},x_{1}^{(2)},\ldots
,x_{N/n}^{(2)},\ldots ,x_{1}^{(n)},\ldots ,x_{N/n}^{(n)})\kappa _{1}\kappa
_{2}\cdots \kappa _{n} & \text{for }N/n\text{ odd}%
\end{array}%
\right.  \notag \\
&=&\mathrm{sgn}(x_{1}^{(1)},\ldots ,x_{N/n}^{(1)},x_{1}^{(2)},\ldots
,x_{N/n}^{(2)},\ldots ,x_{1}^{(n)},\ldots ,x_{N/n}^{(n)}).
\end{eqnarray}%
The proof given in Appendix \ref{sec:appFC} shows directly that the state (%
\ref{eq:wfSim}) with $\chi $ given by (\ref{eq:chi}) and $z_{j}$ arbitrary
is an SU($n$) singlet without referring to the SU($n$) HS model.

\subsection{Wave functions in the hardcore boson basis}

In order to compare the state (\ref{eq:Jastrow}) to known models in
particular limits, we shall now express the state in a hardcore boson basis.
In this picture, the coordinates $z_{j}$ are lattice sites that can be empty
or occupied by at most one hardcore boson. A spin in the state $|n\rangle $
is interpreted as an empty site, and a spin in the state $|\alpha \rangle $,
with $\alpha \in \{1,2,\ldots ,n-1\}$, is interpreted as a site occupied by
a hardcore boson with color $\alpha $.

Referring to (\ref{eq:sunstates}), we observe that the $(n-1)$th component $%
m_{\alpha _{j},n-1}$ of the vector $\vec{m}_{\alpha _{j}}$ can be written as
\begin{equation}
m_{\alpha _{j},n-1}=\frac{n}{\sqrt{2n(n-1)}}p_{j}-\frac{n-1}{\sqrt{2n(n-1)}},
\end{equation}%
where $p_{j}$ is one if $\alpha _{j}\in \{1,2,\ldots ,n-1\}$ and zero if $%
\alpha _{j}=n$. In other words, we can use this component to distinguish
between occupied sites and holes, and we shall use this observation to
eliminate the coordinates of the unoccupied sites from the Jastrow factor in
(\ref{eq:Jastrow}). The part of this factor that includes the contribution
from $m_{\alpha _{j},n-1}$ can be written as
\begin{eqnarray}
\prod_{i<j}(z_{i}-z_{j})^{2m_{\alpha _{i},n-1}m_{\alpha _{j},n-1}}
&=&\prod_{i<j}(z_{i}-z_{j})^{2[\frac{n}{\sqrt{2n(n-1)}}p_{i}-\frac{n-1}{%
\sqrt{2n(n-1)}}][\frac{n}{\sqrt{2n(n-1)}}p_{j}-\frac{n-1}{\sqrt{2n(n-1)}}]}
\notag  \label{eq:Jastrow1} \\
&\propto &\prod_{i<j}(z_{i}-z_{j})^{\frac{n}{n-1}p_{i}p_{j}}%
\prod_{i<j}(z_{i}-z_{j})^{-(p_{i}+p_{j})}.
\end{eqnarray}%
The second factor in the above expression can be simplified as \cite%
{nielsen2012,tu2014}%
\begin{equation}
\prod_{i<j}(z_{i}-z_{j})^{-(p_{i}+p_{j})}=\prod_{j}(-1)^{(j-1)p_{j}}%
\prod_{i}[f_{N}(z_{i})]^{p_{i}},
\end{equation}%
where
\begin{equation}
f_{N}(z_{i})=\prod_{j(\neq i)}(z_{i}-z_{j})^{-1}.
\end{equation}%
Let us next consider the part of the Jastrow factor that includes the
contributions from $m_{\alpha _{j},l}$ with $l=1,2,\ldots ,n-2$. Utilizing (%
\ref{eq:sunstates}) and (\ref{eq:mdotm}), we find
\begin{equation}
\sum_{l=1}^{n-2}m_{\alpha ,l}m_{\alpha ^{\prime },l}=\frac{1}{2}\delta
_{\alpha \alpha ^{\prime }}-\frac{1}{2(n-1)}\quad (\alpha \neq n,\;\alpha
^{\prime }\neq n).
\end{equation}%
If $\alpha $ or $\alpha ^{\prime }$ is equal to $n$, we instead get $%
\sum_{l=1}^{n-2}m_{\alpha ,l}m_{\alpha ^{\prime },l}=0$ as follows
immediately from (\ref{eq:sunstates}). The part of the Jastrow factor that
includes the contributions for $m_{\alpha _{j},l}$ with $l=1,2,\ldots ,n-2$
can therefore be written as
\begin{equation}
\prod_{i<j}(z_{i}-z_{j})^{2\sum_{l=1}^{n-2}m_{\alpha _{i},l}m_{\alpha
_{j},l}}=\prod_{i<j}(z_{i}-z_{j})^{(\delta _{\alpha _{i}\alpha _{j}}-\frac{1%
}{n-1})p_{i}p_{j}}.  \label{eq:Jastrow2}
\end{equation}%
Combining (\ref{eq:Jastrow1}) and (\ref{eq:Jastrow2}), we get the expression
\begin{equation}
\prod_{i<j}(z_{i}-z_{j})^{2\vec{m}_{\alpha _{i}}\cdot \vec{m}_{\alpha
_{j}}}\propto \prod_{i<j}(z_{i}-z_{j})^{(\delta _{\alpha _{i}\alpha
_{j}}+1)p_{i}p_{j}}\prod_{j}(-1)^{(j-1)p_{j}}\prod_{j}[f_{N}(z_{j})]^{p_{j}}
\end{equation}%
for the Jastrow factor.

We would like to also remove the hole coordinates from the sign factor $\chi
$. Doing so gives rise to a sign factor that compensates the factor $%
\prod_{j}(-1)^{(j-1)p_{j}}$ in the wave function. The remaining sign factor
is then $\mathrm{sgn}(x_{1}^{(1)},\ldots ,x_{N/n}^{(1)},x_{1}^{(2)},\ldots
,x_{N/n}^{(2)},\ldots ,x_{1}^{(n-1)},\ldots ,x_{N/n}^{(n-1)})$. We note,
however, that this factor can be absorbed by rearranging the ordering in the
Jastrow factor. Putting everything together, we thus conclude that the state
(\ref{eq:wf1}) can also be written as
\begin{equation}
|\Psi _{\mathrm{F}}\rangle =\sum_{\{x^{(1)}\},\{x^{(2)}\},\ldots
,\{x^{(n-1)}\}}\Psi_{\mathrm{F}} (\{x^{(1)}\},\{x^{(2)}\},\ldots
,\{x^{(n-1)}\})|\{x^{(1)}\},\{x^{(2)}\},\ldots ,\{x^{(n-1)}\}\rangle ,
\end{equation}%
where the sum is over all possible distributions of the $(n-1)N/n$ colored
bosons on the $N$ lattice sites with at most one boson per site, and
\begin{equation}
\Psi _{\mathrm{F}}(\{x^{(1)}\},\{x^{(2)}\},\ldots ,\{x^{(n-1)}\})\propto
\prod_{\alpha }\prod_{i<j}(z_{x_{i}^{(\alpha )}}-z_{x_{j}^{(\alpha
)}})^{2}\prod_{\alpha <\beta }\prod_{i,j}(z_{x_{i}^{(\alpha
)}}-z_{x_{j}^{(\beta )}})\prod_{\alpha }\prod_{j}f_{N}(z_{x_{j}^{(\alpha )}})
\label{eq:wfboson}
\end{equation}%
with $\alpha ,\beta \in \{1,2,\ldots ,n-1\}$ and $i,j\in \{1,2,\ldots ,N/n\}$%
. We shall now comment further on (\ref{eq:wfboson}) for particular choices
of the lattice.

\subsubsection{Jastrow wave functions for the uniform 1D lattice}

We first consider a uniform lattice in 1D with periodic boundary conditions,
which is achieved by choosing $z_{j}=e^{2\pi ij/N}$. For this particular
case, we have the simple expression $f_{N}(z_{j}^{(\alpha )})\propto
z_{j}^{(\alpha )}$ \cite{nielsen2011}. Inserting this in (\ref{eq:wfboson}),
we see that the wave function for the particular case of a uniform 1D
lattice reduces to the ground state of the SU($n$) HS Hamiltonian \cite%
{kawakami1992,ha1992,kiwata1992}%
\begin{equation}
\Psi _{\mathrm{HS}}(\{x^{(1)}\},\{x^{(2)}\},\ldots
,\{x^{(n-1)}\})=\prod_{\alpha }\prod_{i<j}(z_{x_{i}^{(\alpha
)}}-z_{x_{j}^{(\alpha )}})^{2}\prod_{\alpha <\beta
}\prod_{i,j}(z_{x_{i}^{(\alpha )}}-z_{x_{j}^{(\beta )}})\prod_{\alpha
}\prod_{j}z_{x_{j}^{(\alpha )}}.
\end{equation}

\subsubsection{2D Halperin wave functions}

Let us next consider a regular lattice in 2D. We shall assume that the area
of each lattice site (defined as the area of the region consisting of all
points that are closer to the given lattice site than to any other lattice
site) is the same for all lattice sites. In this case, it has been shown in
\cite{tu2014} that
\begin{equation}
|f_{N\rightarrow \infty }(z)|\propto e^{-|z|^{2}/4}
\end{equation}%
for $N$ large. The state (\ref{eq:wfboson}) can therefore be written as
\begin{equation}
\Psi _{\mathrm{F}}^{N\rightarrow \infty }(\{x^{(1)}\},\ldots
,\{x^{(n-1)}\})\propto \prod_{\alpha }\prod_{j}e^{-ig_{x_{j}^{(\alpha
)}}}\prod_{\alpha }\prod_{i<j}(z_{x_{i}^{(\alpha )}}-z_{x_{j}^{(\alpha
)}})^{2}\prod_{\alpha <\beta }\prod_{i,j}(z_{x_{i}^{(\alpha
)}}-z_{x_{j}^{(\beta )}})\prod_{\alpha }\prod_{j}e^{-|z_{x_{j}^{(\alpha
)}}|^{2}/4}  \label{eq:2DHalperin}
\end{equation}%
in the thermodynamic limit, where
\begin{equation}
g_{x_{j}^{(\alpha )}}=\mathrm{Im}(\sum_{k(\neq x_{j}^{(\alpha )})}\ln
(z_{x_{j}^{(\alpha )}}-z_{k})).
\end{equation}

Up to a local phase factor that can be removed with a simple transformation
(of both the wave function and the parent Hamiltonian), we thus observe that
the wave function (\ref{eq:wfSim}) reduces to the lattice version of the
Halperin state \cite{halperin1983}, which appeared in the context of the
multilayer FQH effect. For example, the SU(3) state corresponds to
Halperin's 221 double-layer spin-singlet state. One consequence of this
interesting connection is that the wave function (\ref{eq:wfSim}) describes
an SU($n$) chiral spin liquid state, supporting Abelian anyonic excitations
(the same as those in Halperin states). Another consequence is that the
particular series of Halperin FQH states in (\ref{eq:2DHalperin}) have a
hidden enhanced SU($n$) symmetry. For instance, one may expect that the
chiral gapless edge excitations of these states are described by the SU($n$)$%
_{1}$ WZW model.

\subsection{Numerical results}

\label{sec:numFF}

Since the properties of the uniform 1D SU($n$) HS state are already
well-known, we shall here only investigate the states in 2D. We compute the
TEE $-\gamma $ by considering the state on an $R\times L$ square lattice on
the cylinder and using the formula \cite{kitaev2006,levin2006,jiang2012}
\begin{equation}
S_{L}=\xi L-\gamma  \label{eq:SLF}
\end{equation}%
for the entanglement entropy of half of the cylinder. In (\ref{eq:SLF}), we
assume the cut to be perpendicular to the cylinder axis, $L$ is the number
of spins along the cut, and the formula is valid asymptotically for large $L$
and $R$. The mapping of the IDMPS (\ref{eq:wf2}) to a cylinder is done
through a conformal transformation, which amounts to choosing
\begin{equation}
z_{j}=\exp (2\pi (r_{j}+il_{j})/L)
\end{equation}%
and considering $r_{j}$ and $l_{j}$ as the coordinates rather than $\mathrm{%
Re}(z_{j})$ and $\mathrm{Im}(z_{j})$. This will also change the chiral
correlator by a constant factor, but we can ignore this, since the factor
does not depend on the state of the spins. The square lattice is then
obtained by choosing $r_{j}\in \{-R/2+1/2,-R/2+3/2,\ldots ,R/2-1/2\}$ and $%
l_{j}\in \{1,2,\ldots ,L\}$ and $N=RL$.

Since it is easier to compute numerically, we choose to consider the Renyi
entropy with index 2, which is defined as $S_{L}^{(2)}=-\ln (\text{Tr}(\rho
_{L}^{2}))$, where $\rho _{L}$ is the reduced density operator of half of
the cylinder. Let us label the spins in the left half of the cylinder by the
indices $1,2,\ldots ,N/2$. As observed in \cite{cirac2010,hastings2010}, one
can use the Metropolis Monte Carlo algorithm to compute $\exp (-S_{L}^{(2)})$
by noting that
\begin{eqnarray}
\exp (-S_{L}^{(2)}) &=&\sum_{\alpha _{1},\ldots ,\alpha _{N},\alpha
_{1}^{\prime },\ldots ,\alpha _{N}^{\prime }}\frac{\Psi _{\mathrm{F}}(\alpha
_{1}^{\prime },\ldots ,\alpha _{\frac{N}{2}}^{\prime },\alpha _{\frac{N}{2}%
+1},\ldots ,\alpha _{N})\Psi _{\mathrm{F}}(\alpha _{1},\ldots ,\alpha _{%
\frac{N}{2}},\alpha _{\frac{N}{2}+1}^{\prime },\ldots ,\alpha _{N}^{\prime })%
}{\Psi _{\mathrm{F}}(\alpha _{1},\ldots ,\alpha _{N})\Psi _{\mathrm{F}%
}(\alpha _{1}^{\prime },\ldots ,\alpha _{N}^{\prime })}  \notag
\label{eq:MCentropy} \\
&&\times P(\alpha _{1},\ldots ,\alpha _{N},\alpha _{1}^{\prime },\ldots
,\alpha _{N}^{\prime })
\end{eqnarray}%
and interpreting
\begin{equation}
P(\alpha _{1},\ldots ,\alpha _{N},\alpha _{1}^{\prime },\ldots ,\alpha
_{N}^{\prime })=\frac{|\Psi _{\mathrm{F}}(\alpha _{1},\ldots ,\alpha
_{N})|^{2}|\Psi _{\mathrm{F}}(\alpha _{1}^{\prime },\ldots ,\alpha
_{N}^{\prime })|^{2}}{\left( \sum_{\alpha _{1}^{\prime \prime },\ldots
,\alpha _{N}^{\prime \prime }}|\Psi _{\mathrm{F}}(\alpha _{1}^{\prime \prime
},\ldots ,\alpha _{N}^{\prime \prime })|^{2}\right) ^{2}}
\end{equation}%
as a classical probability distribution. The results of the computations are
shown as a function of the number of spins along the cut in Fig.~\ref%
{fig:TEEF}. The figure provides evidence for $n=3$ and $n=4$ that the TEE is
$-\gamma =-\ln (n)/2$. This is consistent with the prediction that the
states in 2D are chiral spin liquid states, with the SU($n$)$_{1}$ WZW model
being their corresponding chiral edge CFT: According to the fusion rule (\ref%
{eq:SUnWZW}) of the SU($n$)$_{1}$ WZW model, the states support $n$ types of
Abelian anyons with quantum dimension $1$, giving rise to a total quantum
dimension $\sqrt{n}$.

\begin{figure}[tbp]
\includegraphics[width=0.5\textwidth]{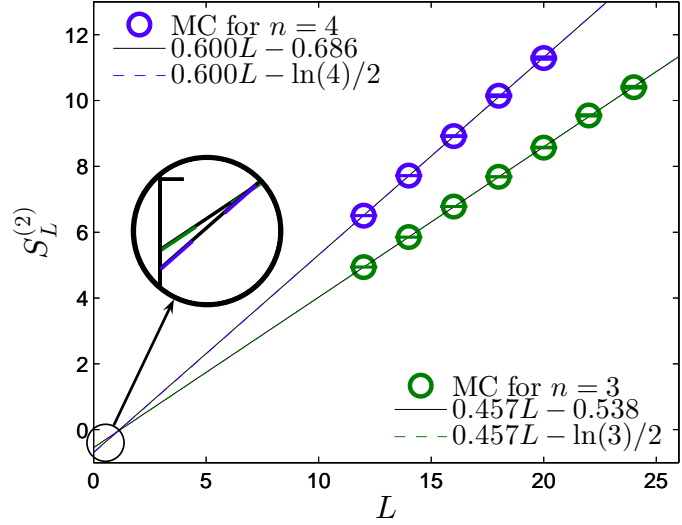}
\caption{Renyi entanglement entropy $S_{L}^{(2)}$ of the 2D IDMPSs (\protect
\ref{eq:wf}) obtained from the fundamental representation of SU($n$) for $%
n=3 $ and $n=4$. The states are defined on an $R\times L$ square lattice on
the cylinder, the cylinder is cut into two halves, and $L$ is the number of
spins along the cut. The length of the cylinder is $R=12$ lattice sites in
both cases. The intersection with the $y$-axis gives the TEE. The points are
obtained from Monte Carlo simulations, and the lines are linear fits with
the constant term being a free parameter (solid lines) or being fixed at $%
-\ln (n)/2$ (dashed lines). The inset is an enlarged view. }
\label{fig:TEEF}
\end{figure}

\section{Parent Hamiltonians for the states from the fundamental
representation}

\label{sec:hamFF}

In this section, we derive parent Hamiltonians of the states $\Psi _{\mathrm{%
F}}$ in Eq.~(\ref{eq:Jastrow}). In 1D, we obtain two-body parent
Hamiltonians, including the SU($n$) HS model as a special case, and for 2D
lattices they are parent Hamiltonians of the SU($n$) chiral spin liquid
states.

Our starting point is the fact that the operator $\mathcal{C}_{i}^{a}$ in (%
\ref{eq:Annihilator2}) annihilates the state (\ref{eq:Jastrow}) as derived
above, $\mathcal{C}_{i}^{a}|\Psi _{\mathrm{F}}\rangle =0$. It follows that
the positive semi-definite Hermitian operator
\begin{equation}
H=\sum_{i}(\mathcal{C}_{i}^{a})^{\dagger }\mathcal{C}_{i}^{a}  \label{eq:h1F}
\end{equation}%
is therefore a parent Hamiltonian of (\ref{eq:Jastrow}), $H|\Psi _{\mathrm{F}%
}\rangle =0$. Inserting (\ref{eq:Annihilator2}) in (\ref{eq:h1F}) and
utilizing the formulas listed in Appendix \ref{sec:appSUn}, we obtain the
explicit expression
\begin{eqnarray}
H &=&\frac{(n-1)(n+2)}{2(n+1)}\sum_{i\neq j}|w_{ij}|^{2}(\vec{t}_{i}\cdot
\vec{t}_{j})+\frac{1}{2}\sum_{i\neq j\neq k}w_{ij}^{\ast }w_{ik}(\vec{t}%
_{j}\cdot \vec{t}_{k})  \notag  \label{eq:HamF} \\
&&+\sum_{i\neq j\neq k}\left( w_{ij}^{\ast }w_{ik}-\frac{1}{n+1}w_{ik}^{\ast
}w_{ij}\right) (\vec{t}_{i}\cdot \vec{t}_{k})(\vec{t}_{i}\cdot \vec{t}_{j})+%
\frac{(n-1)(n+2)}{4n}\sum_{i\neq j}|w_{ij}|^{2},
\end{eqnarray}%
which is valid for general $z_{j}$.

\subsection{Exchange form of the parent Hamiltonian}

As we shall now show, $H$ can also be expressed in terms of the exchange
operator $P_{ij}$, which swaps the spin states at sites $i$ and $j$, i.e., $%
P_{ij}=\sum_{\alpha ,\beta =1}^{n}|\alpha _{i},\beta _{j}\rangle \langle
\beta _{i},\alpha _{j}|$. To do so, we define the following fermionic
representation of the SU($n$) generators%
\begin{equation}
t_{i}^{a}=\sum_{\alpha \beta }c_{i\alpha }^{\dagger }(t^{a})_{\alpha \beta
}c_{i\beta }
\end{equation}%
with the local constraint $\sum_{\alpha =1}^{n}c_{i\alpha }^{\dagger
}c_{i\alpha }=1$ for all $i$. Using Fierz identity (\ref{eq:SUnFierz}), we
can then express the SU($n$) Heisenberg interaction
\begin{eqnarray}
\vec{t}_{i}\cdot \vec{t}_{j} &=&\sum_{\alpha \beta \gamma \delta }c_{i\alpha
}^{\dagger }c_{i\beta }c_{j\gamma }^{\dagger }c_{j\delta }(t^{a})_{\alpha
\beta }(t^{a})_{\gamma \delta }  \notag  \label{eq:Pij} \\
&=&\frac{1}{2}\sum_{\alpha \beta }c_{i\alpha }^{\dagger }c_{i\beta
}c_{j\beta }^{\dagger }c_{j\alpha }-\frac{1}{2n}\sum_{\alpha \gamma
}c_{i\alpha }^{\dagger }c_{i\alpha }c_{j\gamma }^{\dagger }c_{j\gamma }
\notag \\
&=&\frac{1}{2}P_{ij}-\frac{1}{2n}
\end{eqnarray}%
in terms of $P_{ij}$. Inserting this into (\ref{eq:HamF}), we get%
\begin{eqnarray}
H &=&\frac{(n-1)(n+2)}{4(n+1)}\sum_{i\neq j}|w_{ij}|^{2}P_{ij}+\frac{%
(n-1)(n+2)}{4(n+1)}\sum_{i\neq j}|w_{ij}|^{2}-\frac{1}{4(n+1)}\sum_{i\neq
j\neq k}w_{ij}^{\ast }w_{ik}  \notag \\
&&+\frac{1}{4}\sum_{i\neq j\neq k}w_{ij}^{\ast }w_{ik}P_{jk}-\frac{1}{4(n+1)}%
\sum_{i\neq j\neq k}w_{ij}^{\ast }w_{ik}P_{ik}-\frac{1}{4(n+1)}\sum_{i\neq
j\neq k}w_{ij}^{\ast }w_{ik}P_{ij}  \notag \\
&&+\frac{1}{4}\sum_{i\neq j\neq k}w_{ij}^{\ast }w_{ik}P_{ik}P_{ij}-\frac{1}{%
4(n+1)}\sum_{i\neq j\neq k}w_{ij}^{\ast }w_{ik}P_{ij}P_{ik}.
\end{eqnarray}

\subsection{SU($n$) Hamiltonian in 1D}

In this subsection, we restrict ourselves to 1D systems. This is done by
restricting all $z_{j}$ to lie on the unit circle in the complex plane,
i.e.\ $|z_{j}|=1$ $\forall j$. When this is the case, we have $w_{ij}^{\ast
}=-w_{ij}$, and using (\ref{eq:tatb}), the 1D Hamiltonian therefore takes
the form
\begin{eqnarray}
H_{\mathrm{1D}} &=&-\frac{(n-1)(n+2)}{2(n+1)}\sum_{i\neq j}w_{ij}^{2}(\vec{t}%
_{i}\cdot \vec{t}_{j})-\frac{n+2}{2(n+1)}\sum_{i\neq j\neq k}w_{ij}w_{ik}(%
\vec{t}_{j}\cdot \vec{t}_{k})-\frac{(n-1)(n+2)}{4n}\sum_{i\neq j}w_{ij}^{2}
\notag \\
&&-\frac{n}{2(n+1)}\sum_{i\neq j\neq
k}w_{ij}w_{ik}d_{abc}t_{i}^{a}t_{j}^{b}t_{k}^{c}.  \label{eq:1Din}
\end{eqnarray}%
For the particular case $n=2$, the three-body term vanishes because $%
d_{abc}=0$, and we recover the Hamiltonian in Eq.~(70) of \cite{nielsen2011}.

In the following, we simplify (\ref{eq:1Din}). First, by using the cyclic
identity $w_{ij}w_{ik}+w_{ji}w_{jk}+w_{ki}w_{kj}=1$, we find
\begin{equation}
\sum_{i(\neq j,k)}w_{ij}w_{ik}=2w_{jk}^{2}+w_{jk}(c_{j}-c_{k})+(N-2),\quad
c_{j}=\sum_{i(\neq j)}w_{ij},  \label{eq:sumww}
\end{equation}%
and
\begin{equation}
\sum_{i\neq j\neq k}w_{ij}w_{ik}d_{abc}t_{i}^{a}t_{j}^{b}t_{k}^{c}=\frac{1}{3%
}\sum_{i\neq j\neq k}d_{abc}t_{i}^{a}t_{j}^{b}t_{k}^{c}.
\end{equation}%
Inserting these relations into (\ref{eq:1Din}), utilizing that $%
T^{a}=\sum_{i}t_{i}^{a}$ and $w_{ij}^{2}=1+\frac{4z_{i}z_{j}}{%
(z_{i}-z_{j})^{2}}$, the parent Hamiltonian (\ref{eq:1Din}) for the state (%
\ref{eq:Jastrow}) with $|z_{j}|=1$, $\forall j$, can be written as
\begin{eqnarray}
H_{\mathrm{1D}} &=&-2(n+2)\sum_{i\neq j}[\frac{z_{i}z_{j}}{(z_{i}-z_{j})^{2}}%
+\frac{1}{4(n+1)}w_{ij}(c_{i}-c_{j})](\vec{t}_{i}\cdot \vec{t}_{j})  \notag
\\
&&-\frac{n}{6(n+1)}d_{abc}T^{a}T^{b}T^{c}-\frac{n+2}{4(n+1)}%
(2N+n)T^{a}T^{a}-E_{\mathrm{1D}},  \label{eq:1DNonUniformHamiltonian}
\end{eqnarray}%
where $E_{\mathrm{1D}}$ is given by%
\begin{equation}
E_{\mathrm{1D}}=\frac{(n-1)(n+2)}{4n}\sum_{i\neq j}w_{ij}^{2}-\frac{%
(n+2)(n-1)}{12n}N(3N+2n-1).
\end{equation}%
Here let us remind that (\ref{eq:1DNonUniformHamiltonian}) directly comes
from (\ref{eq:h1F}) and $H_{\mathrm{1D}}|\Psi _{\mathrm{F}}\rangle =0$.

Since $|\Psi _{\mathrm{F}}\rangle $ is an SU($n$) singlet, we have $%
d_{abc}T^{a}T^{b}T^{c}|\Psi _{\mathrm{F}}\rangle =T^{a}T^{a}|\Psi _{\mathrm{F%
}}\rangle =0$. Thus, we could get rid of the three-body and two-body
Casimirs in (\ref{eq:1DNonUniformHamiltonian}) and define a pure two-body
parent Hamiltonian%
\begin{equation}
H^{\prime}_{\mathrm{1D}\text{ \textrm{non}}\mathrm{uniform}}=-\sum_{i\neq j}[%
\frac{z_{i}z_{j}}{(z_{i}-z_{j})^{2}}+\frac{1}{4(n+1)}w_{ij}(c_{i}-c_{j})](%
\vec{t}_{i}\cdot \vec{t}_{j}),  \label{eq:1DInhomogeousHamiltonian}
\end{equation}%
which has (\ref{eq:Jastrow}) as its ground state with ground-state energy%
\begin{equation}
E^{\prime}_{\mathrm{1D}\text{ \textrm{non}}\mathrm{uniform}}=\frac{(n-1)}{8n}%
\sum_{i\neq j}w_{ij}^{2}-\frac{n-1}{24n}N(3N+2n-1).
\end{equation}%
The Hamiltonian (\ref{eq:1DInhomogeousHamiltonian}) is an inhomogenous
generalization of the SU($n$) HS model. For $n=2$, it reduces to the SU(2)
inhomogenous HS model derived in \cite{cirac2010}.

\subsection{1D uniform Hamiltonian and the SU($n$) HS model}

\label{sec:1DSUnHS}

We now further restrict $z_{j}$ to be uniformly distributed on the unit
circle by choosing $z_{j}=e^{2\pi ij/N}$. This gives a uniform 1D lattice
with periodic boundary conditions. In this case,
\begin{eqnarray}
c_{j} &=&\sum_{i(\neq j)}w_{ij}=0\text{ \ }\forall j,  \label{eq:UniformSum}
\\
\sum_{i\neq j}w_{ij}^{2} &=&-\frac{1}{3}N(N-1)(N-2).
\label{eq:UniformWijSum}
\end{eqnarray}%
The 1D uniform parent Hamiltonian is therefore
\begin{eqnarray}
H_{\mathrm{1D}\text{ }\mathrm{uniform}} &=&-2(n+2)\sum_{i\neq j}\frac{%
z_{i}z_{j}}{(z_{i}-z_{j})^{2}}(\vec{t}_{i}\cdot \vec{t}_{j})-\frac{n}{6(n+1)}%
d_{abc}T^{a}T^{b}T^{c}  \notag \\
&&-\frac{n+2}{4(n+1)}(2N+n)T^{a}T^{a}-E_{\mathrm{1D}\text{ }\mathrm{uniform}%
},  \label{eq:1Duniform}
\end{eqnarray}%
whose ground-state energy is given by
\begin{equation}
E_{\mathrm{1D}\text{ }\mathrm{uniform}}=-\frac{(n-1)(n+2)}{12n}N(N^{2}+2n+1).
\end{equation}%
We note that the first term in (\ref{eq:1Duniform}) is given by
\begin{equation}
-\sum_{i\neq j}\frac{z_{i}z_{j}}{(z_{i}-z_{j})^{2}}(\vec{t}_{i}\cdot \vec{t}%
_{j})=\sum_{i\neq j}\frac{1}{4\sin ^{2}\frac{\pi }{N}(i-j)}(\vec{t}_{i}\cdot
\vec{t}_{j})=\frac{1}{2}(\frac{N}{\pi })^{2}H_{\mathrm{HS}},  \label{eq:zHS}
\end{equation}%
where $H_{\mathrm{HS}}$ is the 1D SU($n$) HS model%
\begin{equation}
H_{\mathrm{HS}}=\sum_{i<j}\frac{\vec{t}_{i}\cdot \vec{t}_{j}}{(\frac{N}{\pi }%
)^{2}\sin ^{2}\frac{\pi }{N}(i-j)},  \label{eq:SUnHS}
\end{equation}%
In the thermodynamic limit $N\rightarrow \infty $, we have $(\frac{N}{\pi }%
)^{2}\sin ^{2}\frac{\pi }{N}(i-j)\rightarrow 1/(i-j)^{2}$ and the strength
of SU($n$) exchange interaction in (\ref{eq:SUnHS}) is inversely
proportional to the square of the distance between the spins.

Then, we can write the uniform 1D parent Hamiltonian as
\begin{equation}
H_{\mathrm{1D}\text{ }\mathrm{uniform}}=(n+2)(\frac{N}{\pi })^{2}H_{\mathrm{%
HS}}-\frac{n}{6(n+1)}d_{abc}T^{a}T^{b}T^{c}-\frac{n+2}{4(n+1)}%
(2N+n)T^{a}T^{a}-E_{\mathrm{1D}\text{ }\mathrm{uniform}}.
\label{eq:1DUniformHamiltonian}
\end{equation}%
Since $T^{a}|\Psi _{\mathrm{F}}\rangle =0$, the ground-state energy of $H_{%
\mathrm{HS}}$ is given by%
\begin{equation}
E_{\mathrm{HS}}=\frac{1}{n+2}(\frac{\pi }{N})^{2}E_{\mathrm{1D}\text{ }%
\mathrm{uniform}}=-\frac{n-1}{12n}\pi ^{2}(N+\frac{2n+1}{N}).
\label{eq:HSGSenergy}
\end{equation}
The 1D uniform parent Hamiltonian thus practically reduces to the 1D SU($n$)
HS model.

\subsubsection{Energy spectra of the SU($n$) HS model}

For the SU($n$) HS model, it has been shown \cite{haldane1992} that it has a
hidden Yangian symmetry, generated by the total spin operator $T^{a}$ and
the operator%
\begin{equation}
\Lambda ^{a}=\frac{i}{2}\sum_{i\neq j}w_{ij}f_{abc}t_{i}^{b}t_{j}^{c}.
\label{eq:Yangian}
\end{equation}%
We note that $\Lambda ^{a}=\frac{n+1}{n+2}\sum_{i}\mathcal{C}_{i}^{a}$,
which thus annihilates $|\Psi _{\mathrm{F}}\rangle $ as well. It is known\
\cite{haldane1992}\ that $T^{a}$ and $\Lambda ^{b}$ both commute with $H_{%
\mathrm{HS}}$, but they do not mutually commute, which is responsible for
the huge degeneracies in the spectra of $H_{\mathrm{HS}}$.

The eigenvalues of the SU($n$) HS model have been obtained in \cite%
{haldane1992}. Combining (\ref{eq:Pij}) and (\ref{eq:zHS}), we rewrite the
SU($n$) HS Hamiltonian as%
\begin{equation}
H_{\mathrm{HS}}=-(\frac{\pi }{N})^{2}\sum_{i\neq j}\frac{z_{i}z_{j}}{%
(z_{i}-z_{j})^{2}}(P_{ij}-\frac{1}{n})=(\frac{\pi }{N})^{2}H_{\mathrm{Haldane%
}}+(\frac{\pi }{N})^{2}\frac{n-1}{2n}\frac{N(N^{2}-1)}{6},
\label{eq:SUnHaldane}
\end{equation}%
where%
\begin{equation}
H_{\mathrm{Haldane}}=-\sum_{i\neq j}\frac{z_{i}z_{j}}{(z_{i}-z_{j})^{2}}%
(P_{ij}-1).
\end{equation}%
It has been shown \cite{haldane1992} that the complete set of eigenvalues of
$H_{\mathrm{Haldane}}$ can be obtained by the simple formula%
\begin{equation}
H_{\mathrm{Haldane}}|\{m_{i}\}\rangle =\sum_{i}\epsilon
(m_{i})|\{m_{i}\}\rangle ,  \label{eq:Energy}
\end{equation}%
where%
\begin{equation}
\epsilon (m_{i})=m_{i}(m_{i}-N).
\end{equation}%
Here $m_{i}$ are distinct integer rapidities satisfying $m_{i}\in \lbrack
0,N]$ $\forall i$. Physically, the sum of these rapidities is proportional to
the lattice momenta of the energy eigenstate $|\{m_{i}\}\rangle $ \cite%
{haldane1992}%
\begin{equation}
P=\frac{2\pi }{N}\sum_{\{m_{i}\}}m_{i}\text{ (mod }2\pi \text{)}.
\label{eq:Momentum}
\end{equation}

According to \cite{haldane1992}, there is a simple rule for finding
physically allowed sets of rapidities
\begin{equation*}
\{m_{i}\}=\{m_{1},m_{2},\ldots ,m_{M}\}
\end{equation*}%
with $m_{1}<m_{2}<\cdots <m_{M}$ and $M$ is an integer satisfying $M\in
\lbrack 0,\frac{n-1}{n}N]$. The rule is that, all possible sets $%
\{m_{1},m_{2},\ldots ,m_{M}\}$ without $n$\ or more consecutive integers are
allowed and correspond to an eigenstate of $H_{\mathrm{Haldane}}$. For
example, the ground state is represented by the sequence%
\begin{equation}
\{m_{i}\}=\{1,2,\ldots ,n-1,n+1,n+2,\ldots 2n-1,2n+1,\ldots ,N-1\}.
\label{eq:GSrapidity}
\end{equation}%
Using (\ref{eq:Energy}) and (\ref{eq:Momentum}), the energy and lattice
momenta of the ground state are therefore given by%
\begin{equation}
E_{\mathrm{Haldane}}=-\frac{n-1}{6n}(N^{3}+nN),
\end{equation}%
and%
\begin{equation}
P_{\mathrm{GS}}=\frac{n-1}{n}N\pi \text{ (mod }2\pi \text{)}=\left\{
\begin{array}{c}
0 \\
\pi  \\
0%
\end{array}%
\right. \left.
\begin{array}{c}
N/n\text{ even} \\
N/n\text{ odd \& }n\text{ even} \\
N/n\text{ odd \& }n\text{ odd}%
\end{array}%
\right. .  \label{eq:GSmomentum}
\end{equation}%
Note that the ground-state energy $E_{\mathrm{Haldane}}$ determined in this
way is consistent with (\ref{eq:HSGSenergy}) by taking into account the
constant term in (\ref{eq:SUnHaldane}).

\subsubsection{Identifying CFT from finite-size spectra}

CFT gives a powerful prediction for the spectra of 1D critical spin chains.
In particular, it is known that the eigenenergies of a critical quantum
chain with $N$ sites and with periodic boundary conditions are given by \cite%
{blote1986,affleck1986a}
\begin{equation}
E=\varepsilon _{\infty }N-\frac{\pi vc}{6N}+\frac{2\pi v}{N}(h+\bar{h}%
+n_{l}+n_{r}),  \label{eq:CFTenergy}
\end{equation}%
where $\varepsilon _{\infty }$ is the ground-state energy per site in the
thermodynamic limit, $v$ is the spin-wave velocity, $c$ is the central
charge, $h$ and $\bar{h}$ are conformal weights of the primary fields, and $%
n_{l}$ and $n_{r}$ are non-negative integers.

For the SU($n$) HS model, the spin-wave velocity and the conformal weights
of the primary fields can be determined directly by the finite-size spectra
obtained from (\ref{eq:Energy}). To show this, we consider the SU($n$) HS
Hamiltonian in (\ref{eq:SUnHaldane}). Let us start with an excitation
defined through the rapidities%
\begin{equation}
\{m_{i}\}=\{2,\ldots ,n-1,n+1,n+2,\ldots 2n-1,2n+1,\ldots ,N-1\},
\end{equation}%
which is obtained by removing the particle ``1'' in the ground-state
configuration (\ref{eq:GSrapidity}). Using (\ref{eq:Energy}) and (\ref%
{eq:Momentum}), we obtain the excitation energy $E^{\prime }$ and the
lattice momentum $P^{\prime }$ of this excitation%
\begin{eqnarray}
E^{\prime } &=&E_{\mathrm{HS}}+(\frac{\pi }{N})^{2}(N-1)  \notag \\
&=&E_{\mathrm{HS}}+\frac{\pi ^{2}}{N}-\mathcal{O}(1/N^{2}),
\end{eqnarray}%
and%
\begin{equation}
P^{\prime }=P_{\mathrm{GS}}-\frac{2\pi }{N}.
\end{equation}%
Comparing to the CFT prediction of the finite-size spectra (\ref%
{eq:CFTenergy}), this excited state corresponds to $h=\bar{h}=n_{r}=0$ and $%
n_{l}=1$. Thus, we obtain the spin-wave velocity%
\begin{equation}
v=\frac{\pi }{2}.  \label{eq:velocity}
\end{equation}%
However, let us note that the central charge $c$ cannot be obtained using (%
\ref{eq:CFTenergy}). The reason is that the SU($n$) HS Hamiltonian has
long-range interactions, which allow an $N$-dependent constant term and the
ground-state energy as a function of $N$ could violate the CFT prediction (%
\ref{eq:CFTenergy}).

Now we consider other excited states of $H_{\mathrm{HS}}$, by shifting the
sequence of ground-state rapidities in (\ref{eq:GSrapidity}) by $a$, with $%
a\in \{1,\ldots ,n-1\}$. The corresponding rapidity sets are given by%
\begin{equation}
\{m_{i}\}=\{1,\ldots ,[a],\ldots \lbrack n+a],\ldots ,[N-n+a],.\ldots ,N\},
\end{equation}%
where $[qn+a]$, with $q=0,\ldots ,\frac{N}{n}-1$, denotes the missing
rapidities in the rapidity set. By using (\ref{eq:Energy}) and (\ref%
{eq:Momentum}), we obtain the excitation energies and lattice momenta of the
corresponding excited states%
\begin{equation}
E_{a}=E_{\mathrm{HS}}+\frac{a(n-a)}{n}\frac{\pi ^{2}}{N},
\end{equation}%
and%
\begin{equation}
P_{a}=P_{\mathrm{GS}}-a\frac{2\pi }{n}.
\end{equation}%
Note that these excitations are gapless in the thermodynamic limit $%
N\rightarrow \infty $. Compared to the CFT prediction (\ref{eq:CFTenergy}),
these excited states correspond to $h_{a}=\bar{h}_{a}$ and $n_{l}=n_{r}=0$.
Comparing with (\ref{eq:CFTenergy}), we obtain the conformal weights $h_{a}=%
\frac{a(n-a)}{2n}$, which correspond to the primary fields $\Lambda _{a}$ of
the SU($n$)$_{1}$ WZW model. This also agrees with the known results \cite%
{haldane1992,schoutens1994,bouwknegt1996} that the SU($n$)$_{1}$ WZW model
describes the low-energy physics of the SU($n$) HS model.

Regarding the excited states of the SU($n$) HS model, one remaining
interesting question is to obtain their explicit form and to relate them
with the rapidity description in (\ref{eq:Energy}). Some of these excited
states have already been obtained in Refs. \cite{schuricht2005,schuricht2006}%
. As a further remark, we note that the gapless excitations at lattice
momenta $P=a\frac{2\pi }{n}$ with $a\in \{1,\ldots ,n\}$ are also known to
exist in the SU($n$) ULS model \cite{uimin1970,lai1974,sutherland1975},
which belongs to the same SU($n$)$_{1}$ WZW universality class \cite%
{affleck1986b,lauchli2006,fuehringer2008,aguado2009}.

\subsection{SU($n$) Hamiltonian in 2D}

In this subsection, we discuss the parent Hamiltonian in 2D. After
multiplying by an overall constant $\frac{2(n+1)}{(n-1)(n+2)}$, the 2D
Hamiltonian in (\ref{eq:HamF}) can be written as
\begin{eqnarray}
H_{\mathrm{2D}} &=&\sum_{i\neq j}\left[ |w_{ij}|^{2}+\left( \sum_{k(\neq
i,j)}w_{ki}^{\ast }w_{kj}\right) \right] (\vec{t}_{i}\cdot \vec{t}_{j})
\notag \\
&&-\frac{1}{n-1}\sum_{i\neq j\neq k}w_{ij}^{\ast }w_{ik}(if_{abc}-\frac{n}{%
n+2}d_{abc})t_{i}^{a}t_{j}^{b}t_{k}^{c}+\frac{n+1}{2n}\sum_{i\neq
j}|w_{ij}|^{2}.  \label{eq:2DHam}
\end{eqnarray}%
Note that this Hamiltonian can be defined on any 2D lattice (both regular or
irregular) and does not rely on a particular lattice geometry.

For SU(2), we have $d_{abc}=0$ and $f_{abc}t_{i}^{a}t_{j}^{b}t_{k}^{c}=\vec{t%
}_{i}\cdot (\vec{t}_{j}\times \vec{t}_{k})$. Then, (\ref{eq:2DHam}) reduces
to the parent Hamiltonian in \cite{nielsen2012} for the $\nu =1/2$ lattice
Laughlin state. This state is also known as the Kalmeyer-Laughlin state \cite%
{kalmeyer1987,kalmeyer1989}, whose parent Hamiltonian has been extensively
studied \cite%
{laughlin1989,schroeter2004,schroeter2007,thomale2009,kapit2010,greiter2011,nielsen2012,greiter2014}%
. From the parent Hamiltonian, it becomes transparent that the chiral
three-spin interaction term $\vec{t}_{i}\cdot (\vec{t}_{j}\times \vec{t}%
_{k}) $, which explicitly breaks time-reversal and parity symmetries,
stabilizes the spin-1/2 Kalmeyer-Laughlin state. Recently, it has been found
\cite{nielsen2013,bauer2013,bauer2014} that Hamiltonians with short-range
chiral three-spin interactions can already stabilize the Kalmeyer-Laughlin
state. This is very encouraging, as such short-range Hamiltonian might be
realized in cold atomic systems in optical lattices \cite%
{nielsen2013,nielsen2014a}.

The SU($n$) parent Hamiltonian (\ref{eq:2DHam}) also has three-body
interactions. Compared to the SU(2) case, one remarkable feature is that,
the three-body coupling is suppressed by a factor of $1/(n-1)$. This gives
us a hint that, for large $n$, one may have a chance to drop the three-body
terms and the (long-range) Hamiltonian with two-body Heisenberg interactions
may stabilize the lattice Halperin state (\ref{eq:2DHalperin}) as its ground
state. However, as the number of terms in the three-body interactions $%
d_{abc} t^a_i t^b_j t^c_k$ and $f_{abc} t^a_i t^b_j t^c_k$ also increases
with $n$, it is unclear whether the parent Hamiltonian can be adiabatically
connected to the long-range Heisenberg model without closing the gap.
Clarifying whether the gap closes in this interpolation is an interesting
problem and certainly deserves further investigation.

Finally, for the 2D SU($n$) Heisenberg model on a square lattice with only
nearest-neighbor interactions, it has been argued \cite%
{hermele2009,hermele2011} that chiral spin liquid supporting Abelian anyons
becomes stable in the large $n$ limit. Thus, it would be interesting to
further explore its possible connection with our wave function (\ref%
{eq:Jastrow}).

\section{Quantum states from the fundamental and conjugate representations
of SU($n$)}

\label{sec:FC}

In this section, we turn to the more general situation, where we use both
the fundamental and the conjugate representation to construct IDMPSs. In
this case, the chiral correlator (\ref{eq:wavefunction1}) evaluates to
\begin{equation}
\Psi (\alpha _{1},\alpha _{2},\ldots ,\alpha _{N+\bar{N}})=\chi (\alpha
_{1},\alpha _{2},\ldots ,\alpha _{N+\bar{N}})\delta _{\sum_{i=1}^{N+\bar{N}%
}r_{i}\vec{m}_{\alpha _{i}}=0}\prod_{i<j}^{N+\bar{N}%
}(z_{i}-z_{j})^{2r_{i}r_{j}\vec{m}_{\alpha _{i}}\cdot \vec{m}_{\alpha _{j}}},
\end{equation}%
where $\chi (\alpha _{1},\alpha _{2},\ldots ,\alpha _{N+\bar{N}})=\kappa
_{\alpha _{1}}\kappa _{\alpha _{2}}\cdots \kappa _{\alpha _{N+\bar{N}}}$ is
a $z_{j}$-independent phase factor that we shall determine below and
\begin{equation}
r_{j}=\left\{
\begin{array}{cl}
+1 & \text{for }j\in A \\
-1 & \text{for }j\in B%
\end{array}%
\right. .
\end{equation}%
By using (\ref{eq:mdotm}), we can also express the chiral correlator in the
simpler form
\begin{equation}
\Psi (\alpha _{1},\alpha _{2},\ldots ,\alpha _{N+\bar{N}})\propto \chi
(\alpha _{1},\alpha _{2},\ldots ,\alpha _{N+\bar{N}})\delta _{\sum_{i=1}^{N+%
\bar{N}}r_{i}\vec{m}_{\alpha
_{i}}=0}\prod_{i<j}(z_{i}-z_{j})^{r_{i}r_{j}\delta _{\alpha _{i}\alpha
_{j}}}.  \label{eq:wfFCsim}
\end{equation}%
Considering (\ref{eq:sunstates}), we observe that the charge neutrality
condition $\delta _{\sum_{i=1}^{N+\bar{N}}r_{i}\vec{m}_{\alpha _{i}}=0}$
yields%
\begin{eqnarray}
N_{1}-N_{2} &=&N_{\bar{1}}-N_{\bar{2}}  \label{eq:N1N2} \\
N_{1}+N_{2}-2N_{3} &=&N_{\bar{1}}+N_{\bar{2}}-2N_{\bar{3}} \\
&\vdots &  \notag \\
N_{1}+N_{2}+\cdots -(n-1)N_{n} &=&N_{\bar{1}}+N_{\bar{2}}+\cdots -(n-1)N_{%
\bar{n}},
\end{eqnarray}%
where $N_{\alpha }$ ($N_{\bar{\alpha}}$) is the number of spins in the
fundamental (conjugate) representation in the state $|\alpha \rangle $.
Together with the conditions%
\begin{eqnarray}
N_{1}+N_{2}+\cdots +N_{n} &=&N, \\
N_{\bar{1}}+N_{\bar{2}}+\cdots +N_{\bar{n}} &=&\bar{N},
\end{eqnarray}%
we thus conclude that%
\begin{equation}
N_{1}=N_{\bar{1}}+\frac{N-\bar{N}}{n}\text{ \ \ \ \ \ }N_{2}=N_{\bar{2}}+%
\frac{N-\bar{N}}{n}\text{ \ \ \ \ \ }\cdots \text{ \ \ \ \ \ }N_{n}=N_{\bar{n%
}}+\frac{N-\bar{N}}{n}
\end{equation}%
must hold for all nonzero terms in the wave function. This is consistent
with our previous observation that $(N-\bar{N})/n$ must be an integer.

$\chi $ is determined from the requirement that $|\Psi \rangle $ must be a
singlet state, i.e.\ $T^{a}|\Psi \rangle =0$, where $T^{a}=\sum_{i=1}^{N+%
\bar{N}}t_{i}^{a}$. We show explicitly in Appendix \ref{sec:appFC} that this
condition is fulfilled for
\begin{equation}
\chi (\alpha _{1},\alpha _{2},\ldots ,\alpha _{N+\bar{N}})=\mathrm{sgn}%
(x_{1}^{(1,\bar{1})},\ldots ,x_{N_{1}+N_{\bar{1}}}^{(1,\bar{1})},x_{1}^{(2,%
\bar{2})},\ldots ,x_{N_{2}+N_{\bar{2}}}^{(2,\bar{2})},\ldots ,x_{1}^{(n,\bar{%
n})},\ldots ,x_{N_{n}+N_{\bar{n}}}^{(n,\bar{n})}),  \label{eq:chiFC}
\end{equation}%
where $x_{i}^{(\alpha ,\bar{\alpha})}$ is the position within the ket of the
$i$th spin that is in the state $|\alpha \rangle $ without distinguishing
between the fundamental and the conjugate representation. As for the case,
where only the fundamental representation is used, we can obtain (\ref%
{eq:chiFC}) by demanding $\kappa _{\alpha }$ in (\ref{eq:iMPSPrimaryField})
to be Klein factors.

\subsection{Numerical results}

\label{sec:numFC}

\begin{figure}[tbp]
\includegraphics[width=0.5\textwidth]{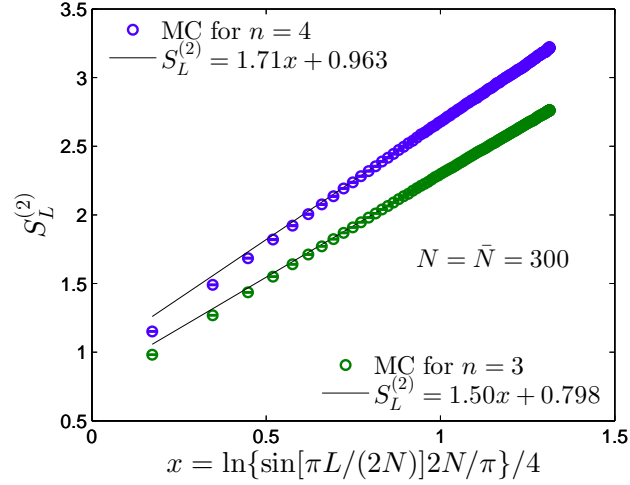}
\caption{Plot of the entanglement entropy of blocks of $L$ consecutive spins
for $N=\bar{N}=300$ and $n=3$ and $n=4$ in 1D. Note that we plot results
only for $L$ even, since the state is only translationally invariant under
translation by an even number of lattice sites. The points are obtained from
Monte Carlo simulations, and the solid lines are linear fits. In computing
the fits, we ignore the 10 leftmost points, since the fits are only
expected to be valid for long distances.}
\label{fig:ent}
\end{figure}

We next investigate the states numerically for lattices with alternating
fundamental and conjugate representations. We start with the uniform 1D
case, where we use the fundamental representation on all the odd sites and
the conjugate representation on all the even sites (see Fig.~\ref%
{fig:FClattice}). Let us consider the entanglement entropy of a block of $L$
consecutive spins, where $L$ is even. We compute this quantity by Monte
Carlo simulations as explained in Sec.~\ref{sec:numFF}, and the result is
shown in Fig.~\ref{fig:ent}. We observe that the entanglement entropy grows
logarithmically. The CFT prediction for the entanglement entropy of a
critical 1D system is \cite{holzhey1994,vidal2003,calabrese2004}
\begin{equation}
S_{L}^{(2)}=\frac{c}{4}\ln \left[ \sin \left( \frac{\pi L}{N+\bar{N}}\right)
\frac{N+\bar{N}}{\pi }\right] +\text{constant},
\end{equation}%
and by using this formula as a fit, we obtain the central charges $c=1.5$
for $n=3$ and $c=1.7$ for $n=4$, respectively.

\begin{figure}[tbp]
\includegraphics[width=0.5\textwidth]{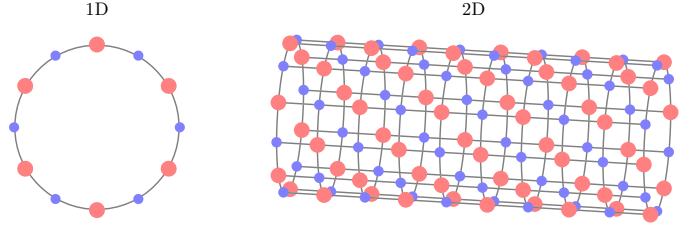}
\caption{Plots of the 1D and 2D lattices with alternating fundamental and
conjugate representations.}
\label{fig:FClattice}
\end{figure}

\begin{figure}[tbp]
\includegraphics[width=0.5\textwidth]{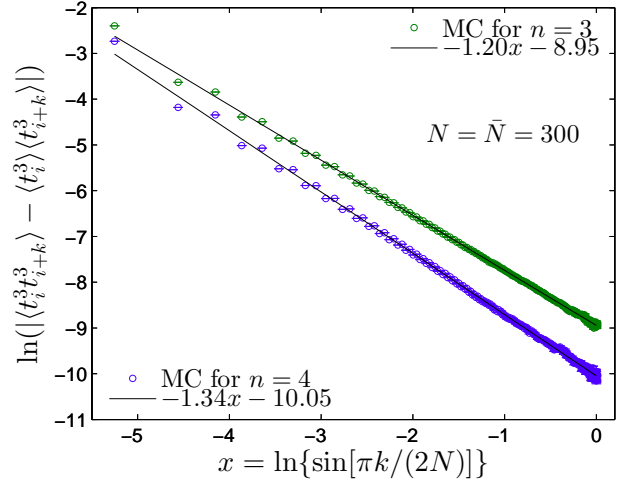}
\caption{Plot of the logarithm of the absolute value of the correlator (%
\protect\ref{eq:cor}) for $N=\bar{N}=300$ and $n=3$ and $n=4$ in 1D. The
sign of the correlator is $(-1)^{k}$. The points are obtained from Monte
Carlo simulations, and the solid lines are linear fits. In computing the
fits, we ignore the 10 leftmost points, since the fits are only expected to
be valid for long distances.}
\label{fig:cor}
\end{figure}

Next we compute the correlation function
\begin{equation}  \label{eq:cor}
c(k)=\langle t^3_{i}t^3_{i+k}\rangle -\langle t^3_{i}\rangle\langle
t^3_{i+k}\rangle
\end{equation}
by using the Metropolis Monte Carlo algorithm. Here, $t^3$ is the third SU($%
n $) generator, which we choose such that $(t^3)_{11}=-(t^3)_{22}=1/2$ in
the fundamental representation and $(t^3)_{11}=-(t^3)_{22}=-1/2$ in the
conjugate representation, whereas all other matrix elements of $t^3$ are
zero in both the fundamental and the conjugate representation (see Appendix %
\ref{sec:appSUn}). The correlator is plotted in Fig.~\ref{fig:cor} for $n=3$
and $n=4$. It is seen to decay algebraically as a function of the chord
distance $\sin[\pi k/(N+\bar{N})]$ with an exponent that is $-1.20$ and $%
-1.34$, respectively.

The logarithmic growth of entanglement entropy and powerlaw decaying
correlation functions both suggest that the 1D state (\ref{eq:wfFCsim}) with
alternating fundamental and conjugate representations describes a critical
spin chain. However, the numerical estimations of the central charges for $%
n=3$ and $4$ show a clear deviation from the SU($n$)$_{1}$ WZW model with $%
c=n-1$. The numerically estimated critical exponents of the two-point
correlation function also differ from $2(n-1)/n$, the expected value for
critical spin chains described by the SU($n$)$_{1}$ WZW model. One
possibility for these deviations is that the system is still described by
the SU($n$)$_{1}$ WZW model, but in the presence of marginally irrelevant
perturbations. Another possibility is that the system belongs to another
universality class which is sharply different from the SU($n$)$_{1}$ WZW
model. In the present framework, it is rather difficult to distinguish these
possibilities. In Sec.~\ref{sec:J2J3chain}, we propose a short-range
Hamiltonian where critical ground states belonging to the same universality
class are likely to appear and which is easier to analyze in practice and
may shed light on the correct critical theory. Another integrable $%
U_{q}[sl(2|1)]$ superspin chain with alternating representations $\mathbf{3}$
and $\mathbf{\bar{3}}$ has been studied in \cite{frahm2012}, which exhibits
several critical theories depending on the parameters of the Hamiltonian.
There could be a connection between these results and our results.

\begin{figure}[tbp]
\includegraphics[width=0.5\textwidth]{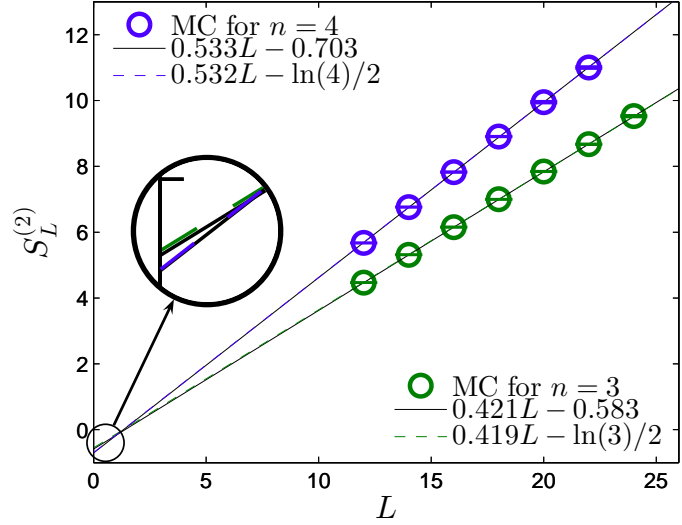}
\caption{Renyi entanglement entropy $S_L^{(2)}$ of the 2D IDMPSs (\protect
\ref{eq:wf}) obtained from the fundamental and conjugate representations of
SU($n$) for $n=3$ and $n=4$. The states are defined on an $R\times L$ square
lattice on the cylinder, the cylinder is cut into two halves in the
direction perpendicular to the axis of the cylinder, and $L$ is the number
of spins along the cut. The length of the cylinder is $R=12$ lattice sites,
and we use the fundamental and conjugate representation on every second site
in a checkerboard pattern. The intersection with the $y$-axis gives the TEE.
The points are obtained from Monte Carlo simulations, and the lines are
linear fits with the constant term being a free parameter (solid lines) or
being fixed at $-\ln(n)/2$ (dashed lines). The inset is an enlarged view. }
\label{fig:TEEFC}
\end{figure}

We now turn to the 2D state on a square lattice on the cylinder, with
fundamental and conjugate representations in a checkerboard pattern (see
Fig.~\ref{fig:FClattice}). In Fig.~\ref{fig:TEEFC}, we compute the TEE
following the same approach as in Sec.~\ref{sec:numFF}. The results are in
agreement with $-\gamma =-\ln (n)/2$ within the precision of the
computation. Similar to the SU($n$) state with only fundamental
representations, this indicates that the states (\ref{eq:wfFCsim}) in 2D are
chiral spin liquids and have the SU($n$)$_{1}$ WZW model as their chiral
edge CFT.

\section{Parent Hamiltonians for the states from the fundamental and
conjugate representations}

\label{sec:hamFC}

As for the case where only the fundamental representation is used, we can
construct a positive semi-definite parent Hamiltonian $H=\sum_{i}(\mathcal{C}%
_{i}^{a})^{\dagger }\mathcal{C}_{i}^{a}$ of the state $|\Psi \rangle $ in (%
\ref{eq:wf}) from the operator (\ref{eq:Annihilator2}) with the property $%
H|\Psi \rangle =0$. Utilizing the formulas listed in Appendix \ref%
{sec:appSUn}, we obtain
\begin{eqnarray}
H &=&\frac{n(n+2)}{4(n+1)}\sum_{i\neq j}|w_{ij}|^{2}(1+r_{i}r_{j}\frac{n-2}{n%
})(\vec{t}_{i}\cdot \vec{t}_{j})+\frac{1}{2}\sum_{i\neq j\neq k}w_{ij}^{\ast
}w_{ik}(\vec{t}_{j}\cdot \vec{t}_{k})  \notag  \label{eq:HamFC} \\
&&+\sum_{i\neq j\neq k}\left( w_{ij}^{\ast }w_{ik}-\frac{1}{n+1}w_{ik}^{\ast
}w_{ij}\right) (\vec{t}_{i}\cdot \vec{t}_{k})(\vec{t}_{i}\cdot \vec{t}_{j})+%
\frac{(n-1)(n+2)}{4n}\sum_{i\neq j}|w_{ij}|^{2},
\end{eqnarray}%
which is valid for general $z_{j}$. Note that this reduces to our previous
result (\ref{eq:HamF}) for $r_{j}=+1$ $\forall j$. We also observe that (\ref%
{eq:HamFC}) does not depend on $r_{j}$ for $n=2$. This happens because the
fundamental representation and the conjugate representation are the same
representation for $n=2$.

\subsection{1D parent Hamiltonian}

We now specialize to 1D by forcing all $z_{j}$ to fulfill $|z_{j}|=1$. This
gives $w_{ij}^{\ast }=-w_{ij}$. We therefore obtain the 1D parent
Hamiltonian
\begin{eqnarray}
H_{\text{1D}} &=&-\frac{n(n+2)}{4(n+1)}\sum_{i\neq j}w_{ij}^{2}(1+r_{i}r_{j}%
\frac{n-2}{n})(\vec{t}_{i}\cdot \vec{t}_{j})-\frac{1}{2}\sum_{i\neq j\neq
k}w_{ij}w_{ik}(\vec{t}_{j}\cdot \vec{t}_{k})  \notag \\
&&-\frac{n}{n+1}\sum_{i\neq j\neq k}w_{ij}w_{ik}(\vec{t}_{i}\cdot \vec{t}%
_{j})(\vec{t}_{i}\cdot \vec{t}_{k})-\frac{(n-1)(n+2)}{4n}\sum_{i\neq
j}w_{ij}^{2}.
\end{eqnarray}%
By using (\ref{eq:tatb}) and (\ref{eq:tatbC}), we find
\begin{equation}
\sum_{i\neq j\neq k}w_{ij}w_{ik}(\vec{t}_{i}\cdot \vec{t}_{j})(\vec{t}%
_{i}\cdot \vec{t}_{k})=\frac{1}{2n}\sum_{i\neq j\neq k}w_{ij}w_{ik}(\vec{t}%
_{j}\cdot \vec{t}_{k})+\frac{1}{2}\sum_{i\neq j\neq
k}w_{ij}w_{ik}r_{i}d_{abc}t_{i}^{a}t_{j}^{b}t_{k}^{c},
\end{equation}%
and by using (\ref{eq:sumww}) and the definition of $T^{a}$, we get
\begin{equation}
\sum_{i\neq j\neq k}w_{ij}w_{ik}(\vec{t}_{j}\cdot \vec{t}_{k})=\sum_{i\neq
j}[2w_{ij}^{2}+w_{ij}(c_{i}-c_{j})](\vec{t}_{i}\cdot \vec{t}_{j})+(N_{\text{T%
}}-2)T^{a}T^{a}-\frac{n^{2}-1}{2n}N_{\text{T}}(N_{\text{T}}-2).
\end{equation}%
Inserting these expressions in the expression for the Hamiltonian leads to
\begin{eqnarray}
H_{\text{1D}} &=&-\frac{(n+2)}{2(n+1)}\sum_{i\neq j}[w_{ij}^{2}(\frac{n+4}{2}%
+r_{i}r_{j}\frac{n-2}{2})+w_{ij}(c_{i}-c_{j})](\vec{t}_{i}\cdot \vec{t}_{j})
\notag \\
&&-\frac{n}{2(n+1)}\sum_{i\neq j\neq
k}w_{ij}w_{ik}r_{i}d_{abc}t_{i}^{a}t_{j}^{b}t_{k}^{c}-\frac{n+2}{2(n+1)}(N_{%
\text{T}}-2)T^{a}T^{a}-E_{\text{1D}}.
\end{eqnarray}%
where
\begin{equation}
E_{\text{1D}}=\frac{(n-1)(n+2)}{4n}[\sum_{i\neq j}w_{ij}^{2}-N_{\text{T}}(N_{%
\text{T}}-2)].
\end{equation}

\subsection{1D uniform parent Hamiltonian}

For the 1D uniform case, $z_{j}=\exp (i\frac{2\pi }{N_{\text{T}}}j)$. By
using (\ref{eq:UniformSum}) and (\ref{eq:UniformWijSum}), the parent
Hamiltonian therefore simplifies to
\begin{eqnarray}
H_{\text{1D uniform}} &=&-\frac{(n+2)}{4(n+1)}\sum_{i\neq
j}w_{ij}^{2}[n+4+r_{i}r_{j}(n-2)](\vec{t}_{i}\cdot \vec{t}_{j})  \notag
\label{eq:1DuniformH} \\
&&-\frac{n}{2(n+1)}\sum_{i\neq j\neq
k}w_{ij}w_{ik}r_{i}d_{abc}t_{i}^{a}t_{j}^{b}t_{k}^{c}-\frac{n+2}{2(n+1)}(N_{%
\text{T}}-2)T^{a}T^{a}-E_{\text{1D uniform}},
\end{eqnarray}%
where
\begin{equation}
E_{\text{1D uniform}}=-\frac{(n-1)(n+2)}{12n}N_{\text{T}}(N_{\text{T}%
}^{2}-4).
\end{equation}%
We plot examples of spectra of $H_{\text{1D uniform}}$ in Fig.~\ref%
{fig:spectrum}. The spectra show that the ground state is unique.

\begin{figure}[tbp]
\includegraphics[width=0.7\textwidth]{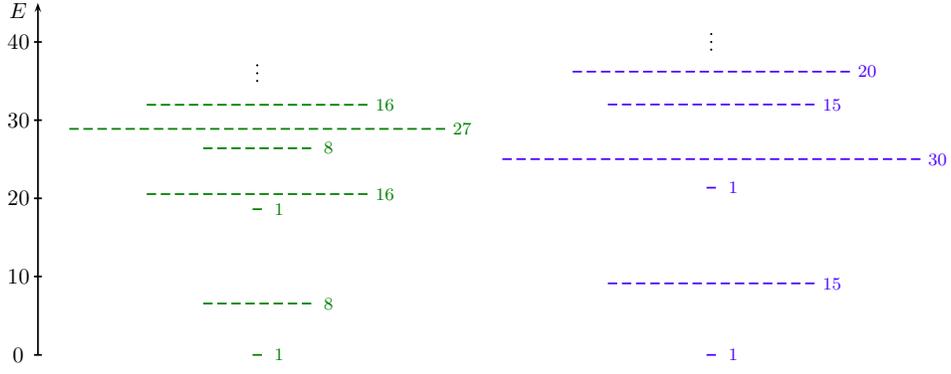}
\caption{Low lying part of the spectrum of $H_{\text{1D uniform}}$ for $N=%
\bar{N}=4$ and fundamental (conjugate) representations on the odd (even)
sites. The results are obtained by exact diagonalization, and the plot on
the left (right) is for $n=3$ ($n=4 $). The numbers written next to the
multiplets are the degeneracies.}
\label{fig:spectrum}
\end{figure}

\subsection{SU($n$) $J_{2}-J_{3}$ chain}

\label{sec:J2J3chain}

One important motivation of studying long-range parent Hamiltonians is that
they may shed light on the physics of some short-range realistic
Hamiltonians. As we already mentioned, the SU($n$) HS Hamiltonian with
inverse-square interactions and the SU($n$) ULS model with only
nearest-neighbor interactions belong to the same SU($n$)$_{1}$ WZW
universality class. For other long-range parent Hamiltonians constructed for
the SU(2)$_{k}$ and SO($n$)$_{1}$ WZW models \cite%
{nielsen2011,thomale2012,paredes2012,tu2013}, the corresponding short-range
Hamiltonians are the SU(2) spin-$\frac{k}{2}$ Takhtajan-Babujian models \cite%
{takhtajan1982,babujian1982} and the SO($n$) Reshetikhin models \cite%
{reshetikhin1985,tu2011}, respectively. Regarding the SU($n$) parent
Hamiltonian (\ref{eq:1DuniformH}) with both fundamental and conjugate
representations, the natural question one may ask is whether there exist
short-range Hamiltonians belonging to the same universality class. In fact,
finding such short-range Hamiltonians can also be very useful for clarifying
the unsolved issue in Sec.~\ref{sec:numFC} on identifying the critical
theory of these models.

To address this problem, we restrict ourselves to the 1D uniform case with
alternating fundamental and conjugate representations (see Fig.~\ref%
{fig:FClattice}). Following the strategy in \cite{nielsen2013}, we truncate
the long-range interactions in (\ref{eq:1DuniformH}) by keeping only
two-body interactions between nearest-neighbor and next-nearest-neighbor
sites, as well as three-body interaction terms among three consecutive
sites. In the thermodynamic limit, $N_{\text{T}}\rightarrow \infty $, this
procedure yields the following Hamiltonian:%
\begin{equation}
H_{\text{\textrm{truncated}}}=\frac{3(n+2)}{n+1}\sum_{i}\vec{t}_{i}\cdot
\vec{t}_{i+1}+\frac{n+2}{4}\sum_{i}\vec{t}_{i}\cdot \vec{t}_{i+2}+\frac{2n}{%
n+1}\sum_{i}r_{i}d_{abc}t_{i}^{a}t_{i+1}^{b}t_{i+2}^{c}.
\end{equation}%
By using (\ref{eq:anticom}) and (\ref{eq:anticom1}), the three-body
interaction term can be rewritten as%
\begin{equation}
\sum_{i}r_{i}d_{abc}t_{i}^{a}t_{i+1}^{b}t_{i+2}^{c}=\frac{1}{n}\sum_{i}\vec{t%
}_{i}\cdot \vec{t}_{i+2}-\sum_{i}[(\vec{t}_{i}\cdot \vec{t}_{i+1})(\vec{t}%
_{i+1}\cdot \vec{t}_{i+2})+\mathrm{h.c.}],
\end{equation}%
and then the truncated Hamiltonian is expressed as%
\begin{equation}
H_{\text{\textrm{truncated}}}=\frac{3(n+2)}{n+1}\sum_{i}\vec{t}_{i}\cdot
\vec{t}_{i+1}+\frac{n^{2}+3n+10}{4(n+1)}\sum_{i}\vec{t}_{i}\cdot \vec{t}%
_{i+2}-\frac{2n}{n+1}\sum_{i}[(\vec{t}_{i}\cdot \vec{t}_{i+1})(\vec{t}%
_{i+1}\cdot \vec{t}_{i+2})+\mathrm{h.c.}].  \label{eq:TruncatedHam}
\end{equation}

There is no guarantee that the truncated Hamiltonian with precisely the
coupling constants in (\ref{eq:TruncatedHam}) has the same physics as the
long-range parent Hamiltonian (\ref{eq:1DuniformH}). However, the form of (%
\ref{eq:TruncatedHam}) suggests that a candidate short-range Hamiltonian
which shares the same physics might be found in the $J_{2}-J_{3}$ SU($n$)
spin chain%
\begin{equation}
H_{J_{2}-J_{3}}=\sum_{i}\vec{t}_{i}\cdot \vec{t}_{i+1}+J_{2}\sum_{i}\vec{t}%
_{i}\cdot \vec{t}_{i+2}+J_{3}\sum_{i}[(\vec{t}_{i}\cdot \vec{t}_{i+1})(\vec{t%
}_{i+1}\cdot \vec{t}_{i+2})+\mathrm{h.c.}]  \label{eq:J2J3Ham}
\end{equation}%
with $J_{2},J_{3}$ being close to the couplings in (\ref{eq:TruncatedHam}).

\begin{figure}[tbp]
\includegraphics[width=0.6\textwidth]{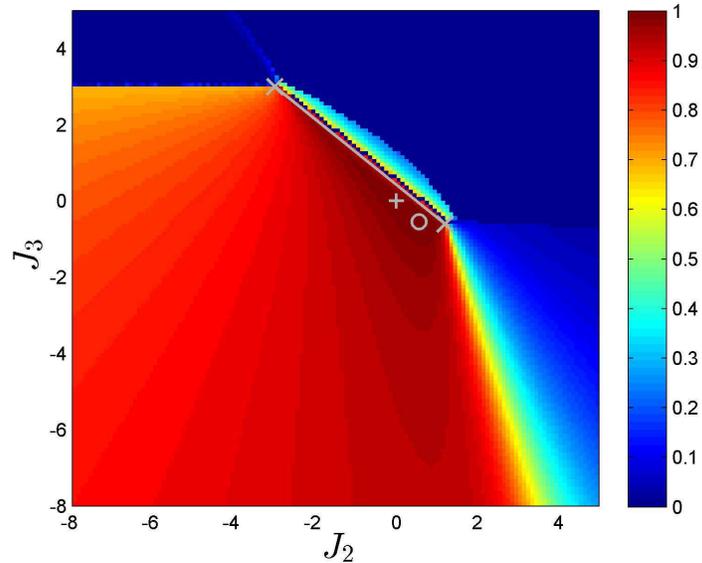}
\caption{Overlap $|\langle \Psi _{J_{2}-J_{3}}|\Psi \rangle |$ between the
ground state of (\protect\ref{eq:J2J3Ham}) and the state (\protect\ref%
{eq:wfFCsim}) as a function of $J_{2}$ and $J_{3}$ for $n=3$ and $N_{\text{T}%
}=10$ sites. The circle denotes the point with $J_{2}=0.557$ and $%
J_{3}=-0.536$, where the maximal overlap $0.9998$ is achieved. The plus sign
corresponds to the pure Heisenberg model with $J_{2}=J_{3}=0$. The
Majumdar-Ghosh model (\protect\ref{eq:OneParameterMG}) is shown by a
straight line terminated at $J_{2}=-3,J_{3}=3$ and $J_{2}=6/5,J_{3}=-3/5$.}
\label{fig:overlap}
\end{figure}

We have performed an exact diagonalization of the Hamiltonian in (\ref%
{eq:J2J3Ham}) for $n=3$ and $N_{\text{T}}=10$ sites. Fig.~\ref{fig:overlap}
shows the overlap $|\langle \Psi _{J_{2}-J_{3}}|\Psi \rangle |$ between the
ground state $|\Psi_{J_{2}-J_{3}}\rangle $ of (\ref{eq:J2J3Ham}) and the
state $|\Psi \rangle $ defined in (\ref{eq:wfFCsim}). The maximum overlap
(marked with a circle in Fig.~\ref{fig:overlap}) is $0.9998$ and occurs for $%
J_{2}=0.557$ and $J_{3}=-0.536$. These values are quite close to $%
J_{2}=0.467 $ and $J_{3}=-0.400$ predicted by the truncated Hamiltonian (\ref%
{eq:TruncatedHam}).

Let us also mention several solvable cases in (\ref{eq:J2J3Ham}), which are
useful for understanding the phase diagram and are also interesting on their
own right. One known solvable point in (\ref{eq:J2J3Ham}) is the pure SU($n$%
)\ Heisenberg chain with $J_{2}=J_{3}=0$, which has gapped dimerized ground
states for $n\geq 3$ \cite{affleck1985,affleck1990}. In Fig. \ref%
{fig:overlap}, this Heisenberg point is marked with a plus sign. Motivated
by a recent work \cite{michaud2012}, we have also identified another class
of solvable cases in (\ref{eq:J2J3Ham}), which have \textit{perfectly}
dimerized ground states and can be viewed as SU($n$) generalizations of the
spin-1/2 Majumdar-Ghosh model \cite{majumdar1969}. These SU($n$)
Majumdar-Ghosh Hamiltonians are written as%
\begin{eqnarray}
H_{\mathrm{MG}} &=&\frac{2}{n}\left( \frac{n-2}{n-1}K_{1}+\frac{n+2}{n+1}%
K_{2}\right) \sum_{i}\vec{t}_{i}\cdot \vec{t}_{i+1}+(K_{2}-K_{1})\sum_{i}%
\vec{t}_{i}\cdot \vec{t}_{i+2}  \notag \\
&&+2\left( \frac{1}{n-1}K_{1}-\frac{1}{n+1}K_{2}\right) \sum_{i}[(\vec{t}%
_{i}\cdot \vec{t}_{i+1})(\vec{t}_{i+1}\cdot \vec{t}_{i+2})+\mathrm{h.c.}],
\label{eq:OneParameterMG}
\end{eqnarray}%
where $K_{1},K_{2}>0$ and which, on a periodic chain with even $N_{\text{T}}$
sites, have ground-state energy $E_{\mathrm{MG}%
}=[(n+1)(n-2)K_{1}+(n-1)(n+2)K_{2}]N_{\text{T}}/(2n^{2})$. In Fig. \ref%
{fig:overlap}, the Majumdar-Ghosh Hamiltonian (\ref{eq:OneParameterMG}) is
shown as a straight line terminated at $J_{2}=-3,J_{3}=3$ and $%
J_{2}=6/5,J_{3}=-3/5$. This line seems to be at a phase boundary between two
different phases. Fully clarifying the phase diagram of (\ref{eq:J2J3Ham})
requires extensive numerics. This is beyond the scope of the present work
and we leave it for a future study.

\section{Conclusion}

\label{sec:conclusion}

In summary, we have constructed a family of spin wave functions with SU($n$)
symmetry from CFT, and we have used the CFT properties of the states to
derive parent Hamiltonians in both 1D and 2D. The states are defined on
arbitrary lattices, and each of the spins transforms under either the
fundamental or the conjugate representation of SU($n$). For the case, where
all spins in the model transform under the fundamental representation, our
results provide a natural generalization of the SU($n$) HS model from a
uniform lattice in 1D to nonuniform lattices in 1D and to 2D. For the
nonuniform 1D case, the Hamiltonian can be chosen to consist of only
two-body terms. In 2D, the states reduce to Halperin type wave functions in
the thermodynamic limit. This suggests that these states are chiral spin
liquids with Abelian anyons, and we find numerically that the total quantum
dimension is close to $\sqrt{n}$. It also shows that a class of Halperin
states have an SU($n$) symmetry and provides parent Hamiltonians that can
stabilize these topological states.

We have also investigated the case with alternating fundamental and
conjugate representations numerically. In 1D, our results suggest that the
state is critical, but the central charges and the exponents of the
correlation functions deviate from the results expected for the SU($n$)$_1$
WZW model. In 2D, we find a nonzero TEE, and the extracted total quantum
dimension is $\sqrt{n}$, which is consistent with the SU($n$)$_1$ WZW model
predictions.

For the case with alternating fundamental and conjugate representations, we
have proposed a short-range Hamiltonian for the 1D uniform case and solved
it exactly for particular choices of the parameters. Given that it is
possible in many related models with long-range Hamiltonians to find
short-range Hamiltonians that describe practically the same low-energy
physics, it is likely that the proposed short-range Hamiltonian has a ground
state in the same universality class as the constructed SU($n$) wave
functions for certain choices of the parameters.

\textit{Note added.}-- During the preparation of this manuscript, we learned
that related results have been obtained by R. Bondesan and T. Quella \cite{bondesan2014}.

\section{Acknowledgment}

The authors acknowledge discussions with Holger Frahm and Kareljan
Schoutens. Our special thanks go to J. Ignacio Cirac for his collaborations
on related topics and for his helpful comments on the present work. This
work has been supported by the EU project SIQS, the DFG cluster of
excellence NIM, FIS2012-33642, QUITEMAD (CAM), and the Severo Ochoa Program.

\appendix

\section{Some useful identities for SU($n$)}

\label{sec:appSUn}

The SU($n$) Lie algebra is formed by $n^{2}-1$ Hermitian and traceless
generators $t^{a}$ ($a=1,\ldots ,n^{2}-1$). They satisfy the commutation
relations%
\begin{equation}
\lbrack t^{a},t^{b}]=if_{abc}t^{c},  \label{eq:com}
\end{equation}%
where $f_{abc}$ is the antisymmetric structure constant of SU($n$). For
SU(2), we have $f_{abc}=\varepsilon _{abc}$.

In the fundamental representation, the generators $t^{a}$ are $n\times n$
matrices that we shall denote by $\tau^{a}$, and in the conjugate
representation the generators are $-(\tau^{a})^*$. For SU(2), a familiar
choice is $\tau^{a}=\frac{1}{2}\sigma ^{a}$, where $\sigma ^{a}$ are Pauli
matrices%
\begin{equation}
\sigma ^{1}=%
\begin{pmatrix}
0 & 1 \\
1 & 0%
\end{pmatrix}%
,\text{ \ }\sigma ^{2}=%
\begin{pmatrix}
0 & -i \\
i & 0%
\end{pmatrix}%
,\text{ \ }\sigma ^{3}=%
\begin{pmatrix}
1 & 0 \\
0 & -1%
\end{pmatrix}%
.
\end{equation}%
For SU(3), it is convenient to define $\tau^{a}=\frac{1}{2}\lambda ^{a}$,
where $\lambda ^{a}$ are the following eight Gell-Mann matrices:%
\begin{eqnarray}
\lambda ^{1} &=&%
\begin{pmatrix}
0 & 1 & 0 \\
1 & 0 & 0 \\
0 & 0 & 0%
\end{pmatrix}%
,\text{ \ }\lambda ^{2}=%
\begin{pmatrix}
0 & -i & 0 \\
i & 0 & 0 \\
0 & 0 & 0%
\end{pmatrix}%
,\text{ \ }\lambda ^{3}=%
\begin{pmatrix}
1 & 0 & 0 \\
0 & -1 & 0 \\
0 & 0 & 0%
\end{pmatrix}%
,\text{ \ }\lambda ^{4}=%
\begin{pmatrix}
0 & 0 & 1 \\
0 & 0 & 0 \\
1 & 0 & 0%
\end{pmatrix}%
,  \notag \\
\lambda ^{5} &=&%
\begin{pmatrix}
0 & 0 & -i \\
0 & 0 & 0 \\
i & 0 & 0%
\end{pmatrix}%
,\text{ \ }\lambda ^{6}=%
\begin{pmatrix}
0 & 0 & 0 \\
0 & 0 & 1 \\
0 & 1 & 0%
\end{pmatrix}%
,\text{ \ }\lambda ^{7}=%
\begin{pmatrix}
0 & 0 & 0 \\
0 & 0 & -i \\
0 & i & 0%
\end{pmatrix}%
,\text{ \ }\lambda ^{8}=\frac{1}{\sqrt{3}}%
\begin{pmatrix}
1 & 0 & 0 \\
0 & 1 & 0 \\
0 & 0 & -2%
\end{pmatrix}%
.
\end{eqnarray}%
The SU(3) Gell-Mann matrices can be straightforwardly generalized to SU($n$)
\cite{georgi1999}. In our present work, we normalize the SU($n$) generators $%
t^{a}$ as $\mathrm{tr}(t^{a}t^{b})=\frac{1}{2}\delta_{ab}$.

The SU($n$) generators in the fundamental representation fulfill
\begin{equation}
\{\tau ^{a},\tau ^{b}\}=\frac{1}{n}\delta _{ab}+d_{abc}\tau ^{c},
\label{eq:anticom}
\end{equation}%
where $d_{abc}$ is symmetric in all indices, and from (\ref{eq:com}) and (%
\ref{eq:anticom}), it follows that%
\begin{equation}
\tau ^{a}\tau ^{b}=\frac{1}{2n}\delta _{ab}+\frac{d_{abc}+if_{abc}}{2}\tau
^{c}.  \label{eq:tatb}
\end{equation}%
For the conjugate representation, we have
\begin{equation}
\{(-(\tau ^{a})^{\ast }),(-(\tau ^{b})^{\ast })\}=\frac{1}{n}\delta
_{ab}-d_{abc}(-(\tau ^{c})^{\ast }),  \label{eq:anticom1}
\end{equation}%
and hence
\begin{equation}
(-(\tau ^{a})^{\ast })(-(\tau ^{b})^{\ast })=\frac{1}{2n}\delta _{ab}+\frac{%
-d_{abc}+if_{abc}}{2}(-(\tau ^{c})^{\ast }).  \label{eq:tatbC}
\end{equation}

The Casimir charge for both the fundamental and the conjugate
representations is given by%
\begin{equation}
t^{a}t^{a}=\frac{n^{2}-1}{2n},  \label{eq:CasimirCharge}
\end{equation}%
and the SU($n$) Fierz identity states that%
\begin{equation}
(t^{a})_{\alpha \beta }(t^{a})_{\gamma \delta }=\frac{1}{2}\delta _{\alpha
\delta }\delta _{\beta \gamma }-\frac{1}{2n}\delta _{\alpha \beta }\delta
_{\gamma \delta }.  \label{eq:SUnFierz}
\end{equation}%
The tensors $d_{abc}$ and $f_{abc}$ satisfy%
\begin{eqnarray}
d_{aab} &=&0, \\
d_{abc}d_{abd} &=&\frac{n^{2}-4}{n}\delta _{cd}, \\
d_{abc}d_{abc} &=&\frac{(n^{2}-1)(n^{2}-4)}{n}, \\
f_{abc}f_{abd} &=&n\delta _{cd}.
\end{eqnarray}%
Additionally, their threefold products are given by \cite{macfarlane1968}%
\begin{eqnarray}
f_{gae}f_{ebh}f_{hcg} &=&-\frac{n}{2}f_{abc}, \\
d_{gae}f_{ebh}f_{hcg} &=&-\frac{n}{2}d_{abc}, \\
d_{gae}d_{ebh}f_{hcg} &=&\frac{n^{2}-4}{2n}f_{abc}, \\
d_{gae}d_{ebh}d_{hcg} &=&\frac{n^{2}-12}{2n}d_{abc}.
\end{eqnarray}%
Using the above identities, we find that
\begin{eqnarray}
t^{a}t^{b}t^{a} &=&-\frac{1}{2n}t^{b}, \\
t^{a}t^{b}t^{c}t^{a} &=&-\frac{1}{2n}t^{b}t^{c}+\frac{1}{4}\delta _{bc}, \\
(\vec{t}_{i}\cdot \vec{t}_{j})^{2} &=&\frac{n^{2}-1}{4n^{2}}-\frac{n}{4}%
(1-r_{i}r_{j}\frac{n^{2}-4}{n^{2}})(\vec{t}_{i}\cdot \vec{t}_{j}),\text{ \ \
\ \ \ }(i\neq j), \\
t_{j}^{a}(\vec{t}_{i}\cdot \vec{t}_{j})t_{i}^{a} &=&\frac{n^{2}-1}{4n^{2}}+%
\frac{n}{4}(1+r_{i}r_{j}\frac{n^{2}-4}{n^{2}})(\vec{t}_{i}\cdot \vec{t}_{j}),%
\text{ \ \ \ \ \ }(i\neq j).
\end{eqnarray}%
In the last two equations we consider two copies $\vec{t}_i$ and $\vec{t}_j$
of SU($n$) generators acting on different sites $i$ and $j$, and $r_k$ is $%
+1 $ ($-1$) if $\vec{t}_k$ belongs to the fundamental (conjugate)
representation.

\section{Global singlet condition}

\label{sec:appFC}

In this Appendix, we prove that the state (\ref{eq:wfFCsim}) with $\chi $
given by (\ref{eq:chiFC}) fulfills $T^{a}|\Psi \rangle =0$, where $%
T^{a}=\sum_{j=1}^{N+\bar{N}}t_{j}^{a}$. First we note that the charge
neutrality condition ensures that the wave function is invariant under the
U(1)$^{\otimes (n-1)}$ subgroup of SU($n$). It is then sufficient to prove
that the operator $\sum_{j=1}^{N+\bar{N}}S_{j}^{12}$ annihilates the state,
where $S_{j}^{12}=t_{j}^{1}+it_{j}^{2}$. The first two generators in the
fundamental representation have the form
\begin{equation*}
\tau ^{1}=\frac{1}{2}%
\begin{pmatrix}
0 & 1 &  \\
1 & 0 &  \\
&  & \ddots%
\end{pmatrix}%
,\quad \tau ^{2}=\frac{1}{2}%
\begin{pmatrix}
0 & -i &  \\
i & 0 &  \\
&  & \ddots%
\end{pmatrix}%
,
\end{equation*}%
where all elements that are not shown are zero. For the $A$ sites
(fundamental representation), we therefore have
\begin{equation}
S_{j}^{12}=\tau _{j}^{1}+i\tau _{j}^{2}=%
\begin{pmatrix}
0 & 1 &  \\
0 & 0 &  \\
&  & \ddots%
\end{pmatrix}%
_{j}=|1\rangle \langle 2|_{j},\quad j\in A,
\end{equation}%
and for the $B$ sites (conjugate representation), we have
\begin{equation}
S_{j}^{12}=(-(\tau _{j}^{1})^{\ast })+i(-(\tau _{j}^{2})^{\ast })=%
\begin{pmatrix}
0 & 0 &  \\
-1 & 0 &  \\
&  & \ddots%
\end{pmatrix}%
_{j}=-|2\rangle \langle 1|_{j},\quad j\in B.
\end{equation}%
Altogether,
\begin{equation}
\sum_{j=1}^{N+\bar{N}}S_{j}^{12}=\sum_{j\in A}|1\rangle \langle
2|_{j}-\sum_{j\in B}|2\rangle \langle 1|_{j}.
\end{equation}

Let us define
\begin{equation}
|\Psi ^{\prime }\rangle =\sum_{j=1}^{N+\bar{N}}S_{j}^{12}|\Psi \rangle .
\end{equation}%
The term in $|\Psi ^{\prime }\rangle $ having $N_{1}+1$ spins in the state $%
|1\rangle $ in the fundamental representation, $N_{\bar{1}}$ spins in the
state $|1\rangle $ in the conjugate representation, $N_{2}-1$ spins in the
state $|2\rangle $ in the fundamental representation, and $N_{\bar{2}}$
spins in the state $|2\rangle $ in the conjugate representation at given
positions has coefficient
\begin{eqnarray*}
&&\Psi ^{\prime }(\{x_{1\rightarrow N_{1}+1}^{(1)}\},\{x_{1\rightarrow
N_{2}-1}^{(2)}\},\ldots ,\{x_{1\rightarrow N_{\bar{1}}}^{(\bar{1}%
)}\},\{x_{1\rightarrow N_{\bar{2}}}^{(\bar{2})}\},\ldots ) \\
&=&\sum_{j=1}^{N_{1}+1}\Psi (\{x_{1\rightarrow j-1}^{(1)},x_{j+1\rightarrow
N_{1}+1}^{(1)}\},\{x_{1\rightarrow N_{2}-1}^{(2)},x_{j}^{(1)}\},\ldots
,\{x_{1\rightarrow N_{\bar{1}}}^{(\bar{1})}\},\{x_{1\rightarrow N_{\bar{2}%
}}^{(\bar{2})}\},\ldots ) \\
&&-\sum_{j=1}^{N_{\bar{2}}}\Psi (\{x_{1\rightarrow
N_{1}+1}^{(1)}\},\{x_{1\rightarrow N_{2}-1}^{(2)}\},\ldots
,\{x_{1\rightarrow N_{\bar{1}}}^{(\bar{1})},x_{j}^{(\bar{2}%
)}\},\{x_{1\rightarrow j-1}^{(\bar{2})},x_{j+1\rightarrow N_{\bar{2}}}^{(%
\bar{2})}\},\ldots ),
\end{eqnarray*}%
where $x_{j}^{(\alpha )}$ ($x_{j}^{(\bar{\alpha})}$) is the index of the $j$%
th spin in the state $|\alpha \rangle $ in the fundamental (conjugate)
representation. We define the order operator $O$ as
\begin{equation}
O(z_{j}-z_{k})=\left\{
\begin{array}{cl}
z_{j}-z_{k} & \text{for }j<k \\
0 & \text{for }j=k \\
z_{k}-z_{j} & \text{for }j>k%
\end{array}%
\right. .  \label{eq:order}
\end{equation}%
Note that
\begin{eqnarray}
&&\Psi (\{x_{1\rightarrow j-1}^{(1)},x_{j+1\rightarrow
N_{1}+1}^{(1)}\},\{x_{1\rightarrow N_{2}-1}^{(2)},x_{j}^{(1)}\},\ldots
,\{x_{1\rightarrow N_{\bar{1}}}^{(\bar{1})}\},\{x_{1\rightarrow N_{\bar{2}%
}}^{(\bar{2})}\},\ldots )  \notag \\
&=&\frac{\mathrm{sgn}(x_{1}^{(1,\bar{1})},\ldots \{x_{j}^{(1)}\text{ missing}%
\}\ldots ,x_{N_{1}+N_{\bar{1}}+1}^{(1,\bar{1})},x_{1}^{(2,\bar{2})},\ldots
,x_{j}^{(1)},\ldots ,x_{N_{2}+N_{\bar{2}}-1}^{(2,\bar{2})},\ldots )}{\mathrm{%
sgn}(x_{1}^{(1,\bar{1})},\ldots ,x_{N_{1}+N_{\bar{1}}+1}^{(1,\bar{1}%
)},x_{1}^{(2,\bar{2})},\ldots ,x_{N_{2}+N_{\bar{2}}-1}^{(2,\bar{2})},\ldots )%
}  \notag \\
&&\times \frac{\prod_{i=1}^{N_{2}-1}O(z_{x_{j}^{(1)}}-z_{x_{i}^{(2)}})}{%
\prod_{i=1(i\neq j)}^{N_{1}+1}O(z_{x_{j}^{(1)}}-z_{x_{i}^{(1)}})}\frac{%
\prod_{i=1}^{N_{\bar{1}}}O(z_{x_{j}^{(1)}}-z_{x_{i}^{(\bar{1})}})}{%
\prod_{i=1}^{N_{\bar{2}}}O(z_{x_{j}^{(1)}}-z_{x_{i}^{(\bar{2})}})}\Psi
(\{x_{1\rightarrow N_{1}+1}^{(1)}\},\{x_{1\rightarrow
N_{2}-1}^{(2)}\},\ldots ,\{x_{1\rightarrow N_{\bar{1}}}^{(\bar{1}%
)}\},\{x_{1\rightarrow N_{\bar{2}}}^{(\bar{2})}\},\ldots )  \notag \\
&=&\frac{\mathrm{sgn}(x_{1}^{(1,\bar{1})},\ldots \{x_{j}^{(1)}\text{ missing}%
\}\ldots ,x_{N_{1}+N_{\bar{1}}+1}^{(1,\bar{1})},x_{j}^{(1)},x_{1}^{(2,\bar{2}%
)},\ldots ,x_{N_{2}+N_{\bar{2}}-1}^{(2,\bar{2})},\ldots )}{\mathrm{sgn}%
(x_{1}^{(1,\bar{1})},\ldots ,x_{N_{1}+N_{\bar{1}}+1}^{(1,\bar{1})},x_{1}^{(2,%
\bar{2})},\ldots ,x_{N_{2}+N_{\bar{2}}-1}^{(2,\bar{2})},\ldots )}  \notag \\
&&\times \frac{\prod_{i=1}^{N_{2}-1}(z_{x_{j}^{(1)}}-z_{x_{i}^{(2)}})}{%
\prod_{i=1(i\neq j)}^{N_{1}+1}O(z_{x_{j}^{(1)}}-z_{x_{i}^{(1)}})}\frac{%
\prod_{i=1}^{N_{\bar{1}}}O(z_{x_{j}^{(1)}}-z_{x_{i}^{(\bar{1})}})}{%
\prod_{i=1}^{N_{\bar{2}}}(z_{x_{j}^{(1)}}-z_{x_{i}^{(\bar{2})}})}\Psi
(\{x_{1\rightarrow N_{1}+1}^{(1)}\},\{x_{1\rightarrow
N_{2}-1}^{(2)}\},\ldots ,\{x_{1\rightarrow N_{\bar{1}}}^{(\bar{1}%
)}\},\{x_{1\rightarrow N_{\bar{2}}}^{(\bar{2})}\},\ldots )  \notag \\
&=&\frac{\prod_{i=1}^{N_{2}-1}(z_{x_{j}^{(1)}}-z_{x_{i}^{(2)}})}{%
\prod_{i=1(\neq j)}^{N_{1}+1}(z_{x_{i}^{(1)}}-z_{x_{j}^{(1)}})}\frac{%
\prod_{i=1}^{N_{\bar{1}}}(z_{x_{i}^{(\bar{1})}}-z_{x_{j}^{(1)}})}{%
\prod_{i=1}^{N_{\bar{2}}}(z_{x_{j}^{(1)}}-z_{x_{i}^{(\bar{2})}})}\Psi
(\{x_{1\rightarrow N_{1}+1}^{(1)}\},\{x_{1\rightarrow
N_{2}-1}^{(2)}\},\ldots ,\{x_{1\rightarrow N_{\bar{1}}}^{(\bar{1}%
)}\},\{x_{1\rightarrow N_{\bar{2}}}^{(\bar{2})}\},\ldots ).
\end{eqnarray}%
In a similar way, we find
\begin{eqnarray*}
&&\Psi (\{x_{1\rightarrow N_{1}+1}^{(1)}\},\{x_{1\rightarrow
N_{2}-1}^{(2)}\},\ldots ,\{x_{1\rightarrow N_{\bar{1}}}^{(\bar{1})},x_{j}^{(%
\bar{2})}\},\{x_{1\rightarrow j-1}^{(\bar{2})},x_{j+1\rightarrow N_{\bar{2}%
}}^{(\bar{2})}\},\ldots ) \\
&=&\frac{\prod_{i=1}^{N_{\bar{1}}}(z_{x_{i}^{(\bar{1})}}-z_{x_{j}^{(\bar{2}%
)}})}{\prod_{i=1(\neq j)}^{N_{\bar{2}}}(z_{x_{j}^{(\bar{2})}}-z_{x_{i}^{(%
\bar{2})}})}\frac{\prod_{i=1}^{N_{2}-1}(z_{x_{j}^{(\bar{2}%
)}}-z_{x_{i}^{(2)}})}{\prod_{i=1}^{N_{1}+1}(z_{x_{i}^{(1)}}-z_{x_{j}^{(\bar{2%
})}})}\Psi (\{x_{1\rightarrow N_{1}+1}^{(1)}\},\{x_{1\rightarrow
N_{2}-1}^{(2)}\},\ldots ,\{x_{1\rightarrow N_{\bar{1}}}^{(\bar{1}%
)}\},\{x_{1\rightarrow N_{\bar{2}}}^{(\bar{2})}\},\ldots ).
\end{eqnarray*}%
To prove that $\Psi ^{\prime }(\{x_{1\rightarrow
N_{1}+1}^{(1)}\},\{x_{1\rightarrow N_{2}-1}^{(2)}\},\ldots
,\{x_{1\rightarrow N_{\bar{1}}}^{(\bar{1})}\},\{x_{1\rightarrow N_{\bar{2}%
}}^{(\bar{2})}\},\ldots )$ vanishes, we thus need to proof that%
\begin{equation}
\sum_{j=1}^{N_{1}+1}\frac{%
\prod_{i=1}^{N_{2}-1}(z_{x_{j}^{(1)}}-z_{x_{i}^{(2)}})}{\prod_{i=1(\neq
j)}^{N_{1}+1}(z_{x_{i}^{(1)}}-z_{x_{j}^{(1)}})}\frac{\prod_{i=1}^{N_{\bar{1}%
}}(z_{x_{i}^{(\bar{1})}}-z_{x_{j}^{(1)}})}{\prod_{i=1}^{N_{\bar{2}%
}}(z_{x_{j}^{(1)}}-z_{x_{i}^{(\bar{2})}})}-\sum_{j=1}^{N_{\bar{2}}}\frac{%
\prod_{i=1}^{N_{\bar{1}}}(z_{x_{i}^{(\bar{1})}}-z_{x_{j}^{(\bar{2})}})}{%
\prod_{i=1(\neq j)}^{N_{\bar{2}}}(z_{x_{j}^{(\bar{2})}}-z_{x_{i}^{(\bar{2}%
)}})}\frac{\prod_{i=1}^{N_{2}-1}(z_{x_{j}^{(\bar{2})}}-z_{x_{i}^{(2)}})}{%
\prod_{i=1}^{N_{1}+1}(z_{x_{i}^{(1)}}-z_{x_{j}^{(\bar{2})}})}=0.
\label{eq:PolynomialIdentity}
\end{equation}%
We rewrite the left-hand side (LHS) of Eq.~(\ref{eq:PolynomialIdentity}) into%
\begin{eqnarray}
\mathrm{LHS} &=&(-1)^{N_{\bar{1}}-N_{1}}\sum_{j=1}^{N_{1}+1}\frac{%
\prod_{i=1}^{N_{\bar{1}}}(z_{x_{j}^{(1)}}-z_{x_{i}^{(\bar{1}%
)}})\prod_{i=1}^{N_{2}-1}(z_{x_{j}^{(1)}}-z_{x_{i}^{(2)}})}{\prod_{i=1(\neq
j)}^{N_{1}+1}(z_{x_{j}^{(1)}}-z_{x_{i}^{(1)}})\prod_{i=1}^{N_{\bar{2}%
}}(z_{x_{j}^{(1)}}-z_{x_{i}^{(\bar{2})}})}  \notag \\
&&+(-1)^{N_{\bar{1}}-N_{1}}\sum_{j=1}^{N_{\bar{2}}}\frac{\prod_{i=1}^{N_{%
\bar{1}}}(z_{x_{j}^{(\bar{2})}}-z_{x_{i}^{(\bar{1})}})%
\prod_{i=1}^{N_{2}-1}(z_{x_{j}^{(\bar{2})}}-z_{x_{i}^{(2)}})}{%
\prod_{i=1(\neq j)}^{N_{\bar{2}}}(z_{x_{j}^{(\bar{2})}}-z_{x_{i}^{(\bar{2}%
)}})\prod_{i=1}^{N_{1}+1}(z_{x_{j}^{(\bar{2})}}-z_{x_{i}^{(1)}})}.
\end{eqnarray}%
Let us denote $z_{x_{1}^{(1)}},\ldots ,z_{x_{N_{1}+1}^{(1)}},z_{x_{1}^{(\bar{%
2})}},\ldots ,z_{x_{N_{\bar{2}}}^{(\bar{2})}}$ as $z_{p}$ $(p=1,2,\ldots
,N_{1}+N_{\bar{2}}+1)$ and $z_{x_{1}^{(\bar{1})}},\ldots ,z_{x_{N_{\bar{1}%
}}^{(\bar{1})}},z_{x_{1}^{(2)}},\ldots ,z_{x_{N_{2}-1}^{(2)}}$ as $w_{l}$ $%
(l=1,2,\ldots ,N_{\bar{1}}+N_{2}-1)$. Then
\begin{equation}
\mathrm{LHS}=(-1)^{N_{\bar{1}}-N_{1}}\sum_{p=1}^{N_{1}+N_{\bar{2}}+1}\frac{%
\prod_{l=1}^{N_{\bar{1}}+N_{2}-1}(z_{p}-w_{l})}{\prod_{q=1(\neq
p)}^{N_{1}+N_{\bar{2}}+1}(z_{p}-z_{q})}.  \label{eq:LHS}
\end{equation}%
From (\ref{eq:N1N2}) it follows that $N_{1}+N_{\bar{2}}=N_{\bar{1}}+N_{2}$.
Multiplying out the polynomial in the numerator, we observe that (\ref%
{eq:LHS}) is zero for all choices of $w_l$ if we have
\begin{equation}  \label{eq:poliden}
\sum_{p=1}^{K}\frac{z_{p}^{m}}{\prod_{q=1(\neq p)}^{K}(z_{p}-z_{q})}=0\qquad
\text{for } m=0,1,\ldots ,K-2,
\end{equation}%
where $K=N_{1}+N_{\bar{2}}+1$. To prove (\ref{eq:poliden}), we first
multiply by the nonzero factor $(-1)^{K-1}\prod_{1\leq q<l\leq K}(z_l-z_q)$,
which transforms the left-hand side of (\ref{eq:poliden}) into
\begin{equation}
\sum_{p=1}^{K} (-1)^{p-1}z_{p}^{m}\prod_{q<l(\neq p)}^{K}(z_{l}-z_{q}).
\end{equation}
This expression contains the Vandermonde determinant. Specifically,
\begin{eqnarray}
&&\sum_{p=1}^{K}(-1)^{p-1}z_{p}^{m}\prod_{q<l(\neq p)}^{K}(z_{l}-z_{q})
\notag \\
&=&\sum_{p=1}^{K}(-1)^{p-1}z_{p}^{m}\det
\begin{pmatrix}
1 & \cdots & 1 & 1 & \cdots & 1 \\
z_1 & \cdots & z_{p-1} & z_{p+1} & \cdots & z_K \\
\vdots & \ddots & \vdots & \vdots & \ddots & \vdots \\
z_1^{K-2} & \cdots & z_{p-1}^{K-2} & z_{p+1}^{K-2} & \cdots & z_K^{K-2}%
\end{pmatrix}
\notag \\
&=&\det
\begin{pmatrix}
z_1^m & z_2^m & \cdots & z_K^m \\
1 & 1 & \cdots & 1 \\
z_1 & z_2 & \cdots & z_K \\
\vdots & \vdots & \ddots & \vdots \\
z_1^{K-2} & z_2^{K-2} & \cdots & z_K^{K-2}%
\end{pmatrix}
\notag \\
&=&0,
\end{eqnarray}%
where the last equality follows because the determinant has two identical
rows for all $m\in\{0,1,\ldots ,K-2\}$. This completes the proof of the
singlet property.

\bibliography{sun}

\end{document}